\documentclass[fleqn,usenatbib]{mnras}

\usepackage{newtxtext,newtxmath}

\usepackage[T1]{fontenc}

\DeclareRobustCommand{\VAN}[3]{#2}
\let\VANthebibliography\thebibliography
\def\thebibliography{\DeclareRobustCommand{\VAN}[3]{##3}\VANthebibliography}

\usepackage{graphicx}	
\usepackage{amsmath}	
\usepackage{subcaption}

\title[Eccentric modes in unstratified MHD discs]{Linear and nonlinear eccentric mode evolution in unstratified MHD discs}

\author[E. M. Lynch and J. W. Dewberry]{
Elliot M. Lynch,$^{1}$\thanks{E-mail: elliot.lynch@ens-lyon.fr}
Janosz W. Dewberry,$^{2}$
\\
$^{1}$Univ Lyon, Univ Lyon1, Ens de Lyon, CNRS, Centre de Recherche Astrophysique de Lyon UMR5574, F-69230, Saint-Genis,-Laval, France\\
$^{2}$ Canadian Institute for Theoretical Astrophysics, 60 St. George Street, Toronto, ON M5S 3H8, Canada,\\
}

\date{Accepted XXX. Received YYY; in original form ZZZ}

\pubyear{2023}

\begin{document}
\label{firstpage}
\pagerange{\pageref{firstpage}--\pageref{lastpage}}
\maketitle

\begin{abstract}
In this paper we develop a framework for studying unstratified, magnetised eccentric discs and compute uniformly precessing eccentric modes in a cylindrical annulus which provide convenient initial conditions for numerical simulations. The presence of a magnetic field in an eccentric disc can be described by an effective gas with a modified equation of state. At magnetic field strengths relevant to the magneto-rotational instability the magnetic field has negligible influence on the evolution of the eccentric disc, however the eccentric disc can significantly enhance the magnetic field strength over that in the a circular disc. We verify the suitability of these eccentric disc solutions by carrying out 2D simulations in RAMSES. Our simulated modes (in 2D) follow a similar evolution to the purely hydrodynamical modes, matching theoretical expectations, provided they are adequately resolved. Such solutions will provide equilibrium states for studies of the eccentric magneto-rotational instability and magnetised parametric instability in unstratified discs and are useful for exploring the response of disc turbulence on top of a fluid flow varying on the orbital timescale.
\end{abstract}




\section{Introduction}

Eccentric gaseous discs, where the gas orbits on Keplerian ellipses, are found in a variety of astrophysical contexts. To date there have been many theoretical and numerical studies considering unmagnetised eccentric discs \citep{Ogilvie01,Ogilvie14,Barker16,Wienkers18,Ogilvie19,Pierens20,Dewberry19b}. Recently several studies have considered the behaviour of magnetic fields in eccentric discs. The effect of magnetic stresses on eccentric discs was considered by \citet{Ogilvie01} who developed a turbulent stress model based on the ideal induction equation in an orbital coordinate system. \citet{Ogilvie14} include an magnetohydrodynamic (MHD) form of their eccentric shearing box model. This was used to study the linear phase of the magneto-rotational instability (MRI) by \citet{Chan18}, while \citet{Lynch21} used the formalism to study the effect of a coherent magnetic field on the disc vertical structure. Global simulation of MRI in eccentric discs were performed by \citet{Dewberry20} who found that sufficiently nonlinear eccentric waves can shut off the MRI. \citet{Oyang21} compared the excitation of eccentricity in MRI turbulent discs to viscous, hydrodynamical discs and found that the latter were excited to larger eccentricities. Finally \citet{Chan22} performed a global simulation of the MRI in an elliptical annulus of large ($e = 0.5$) constant eccentricity motivated by the highly eccentric discs found in tidal disruption events.

One challenge for (hydro or MHD) simulations of eccentric discs is the strong differential precession due to pressure forces which arises for arbitrary eccentricity profiles. Notably this occurs for a uniformly eccentric ring which, naively, might be considered the simplest eccentricity profile to simulate. This strong differential precession was a problem encountered by \citet{Chan22} who utilised an elliptical coordinate system to model a disc of uniform eccentricity, which became significantly misaligned from the simulation grid after only 15 outer disc orbits. Strong differential precession quickly generates large pressure gradients as a result of orbital compression, which can be very difficult to resolve numerically, leading to artificial damping of the disc eccentricity. The strength of this differential precession will depend on the magnetic field strength and configuration. The rapid evolution of the disc orbits can potentially lead to transient phenomena that are primarily a consequence of the choice of initial conditions. This makes it difficult to disentangle the effect of the MRI and parametric instability from the evolution of the non-steady initial conditions. It would thus be beneficial to study how the MRI develops on top of a steady, or slowly evolving, eccentric background.

A solution to this problem can be found in the existence of eccentric modes. These are untwisted eccentric discs with a time independent eccentricity profile which undergo uniform (i.e. rigid body) precession as a result of pressure gradients and other non-Keplerian forces. Eccentric modes are thus a particularly suitable setting for numerical simulations. They are also well motivated physically as they often provide a good approximation to the relaxed state of many eccentric discs when excitation and damping processes are considered \citep{Kley08,Miranda17,Teyssandier16,Teyssandier17,Ragusa17}.

In this paper we extend the Hamiltonian eccentric disc theory of \citet{Ogilvie19} to allow for the inclusion of a large scale, structured, magnetic field in an unstratified disc. We calculate modal (uniformly precessing) solutions for these eccentric MHD discs. Such solutions are not intended as a realistic model of a magnetised eccentric disc, owing to the neglect of important 3D effects \citep{Ogilvie01,Ogilvie08,Ogilvie14,Teyssandier16,Ogilvie19} and unrealistic global field structure \citep[see][for the 3D field structure in a circular disc]{Ogilvie97}, however these solutions are intended to provide a convenient setting for numerical simulations of the eccentric MRI. To this end we run 2D MHD simulations in the code RAMSES \citep{Teyssier2002,Fromang2006,Faure2014} using our calculated eccentric mode as an initial condition to test their suitability for numerical calculations. By using an eccentric mode as our initial condition we aim to avoid strong differential precession, due to pressure, destroying the eccentric disc as seen in \citet{Chan22}.

This paper is structured as follows. In Section \ref{disc geometry} we give an overview of eccentric disc geometry and orbital coordinate systems. In Section \ref{Hamiltonian derivation} we extent the Hamiltonian formalism of \citet{Ogilvie19} to unstratified ideal MHD discs, which we use to derive linear theory in Section \ref{linear theory} and compute nonlinear eccentric modes in Section \ref{nonlinear modes}. We compare the eccentric disc theory against 2D MHD simulations in Section \ref{simulations}. Finally, we present our conclusions in Section \ref{conclusion} and a derivation of the magnetic vector potential is given in the appendix to aid with future numerical work.

\section{Eccentric Disc Geometry} \label{disc geometry}

The geometry of an eccentric disc consists of a set of non-intersecting, confocal Keplerian ellipses where the dominant fluid motion consist of the Keplerian motion. These Keplerian orbits slowly evolve due to the effects of pressure gradients and, in MHD discs, magnetic fields. To describe both the geometry, and dynamics, of an eccentric discs it is often convenient to make use of an orbital coordinate system. This is a coordinate system, based on the orbital elements of celestial mechanics, that describes a point in the mid-plane of the disc by an orbit labelling coordinate, specifying the orbit the point lies on, and a coordinate denoting where along that orbit the point lies Typically such an orbital coordinate system will define a time dependant map from some circular reference disc onto the physical eccentric disc, with the dynamics of the eccentric disc being described by the slow evolution of this orbital coordinate system.

We formulate this orbital coordinate system in terms of a Lagrangian map between the reference and physical variables $\boldsymbol{a} \mapsto \boldsymbol{x}$, where $\boldsymbol{a}$ are the orbital coordinates associated with a fluid element and $\boldsymbol{x}$ is the fluid element position vector in Cartesian coordinates.  This Lagrangian map can be thought of as mapping some reference circular state into the physical eccentric disc, similar to \citet{Ogilvie18} and \citet{Ogilvie19}. We denote the Jacobian associated with this Lagrangian map by $J_{i j} = \frac{\partial x_i}{\partial a^j}$ and introduce the notation

\begin{equation}
 J_3 = \det (J_{i j}) = \frac{J}{J^{\circ}} \frac{H}{H^{\circ}} ,
\end{equation}
where $J$ is the Jacobian determinant of the 2D transform and $H$ is the disc scale height and we adopt the convention that the superscript $\cdot^{\circ}$ denotes a quantity in the reference circular disc. Note that this Jacobian is for an orbital coordinate system using the stretched vertical coordinate $\tilde{z}$. In \citet{Ogilvie18} this is approximated by its value at the midplane, which is valid when the disc is sufficiently thin. For the unstratified discs considered here the horizontal and vertical parts of the transform are separable so this approximation is unnecessary.

We can define an orbital coordinate system, following \citet{Ogilvie19}, where the orbits are labelled by the semimajor axis, a, and the position around the orbit are labelled by the eccentric anomaly $E$. The shape of each orbit is controlled by the orbits eccentricity, $e$, and longitude of pericentre $\varpi$. The $(a,E)$ orbital coordinate system are related to the cylindrical radius through

\begin{equation}
r = a (1 - e \cos E) ,
\end{equation}
and to the azimuthal angle, $\phi$, through the true anomaly $f = \phi - \varpi$ which satisfies,

\begin{equation}
\cos f = \frac{\cos E - e}{1 - e \cos E}, \quad \sin f = \frac{\sqrt{1 - e^2} \sin E}{1 - e \cos E} .
\end{equation}
We can extend this coordinate system to 3D by taking the disc midplane as a reference plane and labelling points by their height above/below the midplane, $z$. One can also introduce a stretched vertical coordinate $\tilde{z} = z/H$, where $H$ is some characteristic vertical lengthscale such as the disc thickness or scale height. In unstratified models this is typically taken to be the sonic length $H = c_s \Omega^{-1}$, with $c_s$ the sound speed. 

It will also often be useful to make use of the mean anomaly $M$, which is related to the eccentric anomaly through

\begin{equation}
M = E - e \sin E = n (t - \tau) ,
\end{equation}
where $n = (G M_1/a^3)^{1/2}$ is the mean motion, $M_1$ is the mass of the central object and $\tau$ is the time of pericentre passage. This allows us to define an $(a,M)$ orbital coordinate system where the position around the orbit is now denoted by the mean anomaly. This can be extended to 3D in the same way as the $(a,E)$ coordinates. 

The Jacobian determinant of the $(a,M)$ orbital coordinate system can be expressed as $J = J^{\circ} j$, where $J^{\circ} = a$ and we have introduced the dimensionless Jacobian determinant from \citet{Ogilvie19},

\begin{align}
\begin{split}
 j = (J/J^{\circ}) &= \frac{1 - e (e + a e_a)}{\sqrt{1 - e^2}} - \frac{a e_a \cos E}{\sqrt{1 - e^2}} - a e \varpi_a \sin E  \\
 &= \frac{1 - e (e + a e_a)}{\sqrt{1 - e^2}} [1 - q \cos (E - E_0)]
\end{split}
\end{align}
which is related to the elliptical geometry of the disc and we have introduced the notation that a subscript $a$ denotes a partial derivative with respect to the semimajor axis. As in \citet{Ogilvie19} we have introduced the orbital intersection parameter, q, and $E_0$, the eccentric anomaly at which the maximum orbital compression occurs. These are related to the orbital elements and their derivatives through

\begin{equation}
 q \cos E_0 = \frac{a e_a}{1 - e (e + a e_a)} , \quad q \sin E_0 = \frac{\sqrt{1 -e^2} a e \varpi_a}{1 - e (e + a e_a)} .    
\end{equation}

\section{Derivation of the Eccentric Disc Hamiltonian} \label{Hamiltonian derivation}

To derive the equation governing the evolution of the eccentric orbits we shall start from the Lagrangian formulation of ideal MHD. After performing a vertical integration we exploit a scale separation which occur in ``thin'' discs where the Lagrangian can be separated into an $O(1)$ contribution from the Keplerian terms and $O(\epsilon^2)$ contributions from the internal and magnetic energies. In an unstratified disc $\epsilon$ should be thought of as a characteristic measure of the reciprocal Mach number (or reciprocal Alfv\'{e}n number for strongly magnetised discs), rather than the aspect ratio used in thin disc theory.

The Lagrangian for ideal MHD is \citep[e.g.][]{Ogilvie16}

\begin{equation}
L = \iiint \rho^{\circ} \left[ \frac{1}{2} u^2 - \Phi(\boldsymbol{x}) - \varepsilon (\boldsymbol{a}, J_{3}) - \frac{J_3 B^2 (\boldsymbol{a}, J_{3}, J_{i j})}{2 \mu_0 \rho_0} \right]  d^3 \boldsymbol{a} ,
\label{mhd lagrangian}
\end{equation}
where $u$ is the fluid velocity, $\rho^{\circ}$ is the density in the reference disc, $B^{i}$ is the disc magnetic field, $\varepsilon$ is the specific internal energy and $\boldsymbol{a}$ are the 3D Lagrangian coordinates. \citet{Ogilvie18} developed a fairly general thin disc model based on affine transforms of fluid elements which provides a convenient setting for formulating eccentric disc models. The affine transform for a coplanar, unwarped, disc is

\begin{equation}
\boldsymbol{x} = \bar{\boldsymbol{x}} + H \tilde{z} \hat{\boldsymbol{e}}_z ,
\end{equation}
where $\bar{\boldsymbol{x}}$ are the coordinates in the disc midplane and  $\tilde{z}$ is a stretched vertical coordinate. For an unstratified disc, where $\Phi(\boldsymbol{x})$ is independent of the vertical coordinate, with periodic boundary conditions in the vertical direction, we can take $H = H^{\circ}$ and treat all quantities as independent of $\tilde{z}$. Therefore we have $J_3 = J/J^{\circ}$, $J_{z z} = 1$, $J_{z x} = J_{z y} = 0$ and we can vertically integrate the Lagrangian (Equation \ref{mhd lagrangian}) to obtain,

\begin{equation}
L = \iint \Sigma^{\circ} \left[ \frac{1}{2} u^2 - \Phi(\bar{\boldsymbol{x}}) - \bar{\varepsilon} (\bar{\boldsymbol{a}}, J) - \frac{(J/J^{\circ}) B^2 H^{\circ}}{2 \mu_0 \Sigma^{\circ}} \right]  d^2 \bar{\boldsymbol{a}} ,
\label{2d mhd lagrangian}
\end{equation}
where $\Sigma^{\circ} = \rho^{\circ} H^{\circ}$ is the surface density of the reference disc and $\bar{\boldsymbol{a}}$ are the 2D Lagrangian coordinates. Here overbars denote a 2D (midplane) coordinate system.

To obtain the disc magnetic field we can look for periodic solutions to the induction equation in an eccentric shearing box \citep{Ogilvie14}. These periodic magnetic field solutions were derived by \citet{Lynch21}, and consists of a vertical field with a quasi-toroidal (orbit following) field, 

\begin{equation}
 B^{i} = B_{t 0} (a,z/H) \frac{J^{\circ}}{J} \frac{H^{\circ}}{H} \frac{v_{\rm orbital}^{i}}{n} + B_{z 0} (a) \frac{J^{\circ}}{J} \hat{e}_{z}^{i} . \label{magnetic field solution}
\end{equation}
Here $B_{t 0}$ and $B_{z 0}$ are the toroidal and vertical magnetic fields in the reference circular disc (we have chosen not to use the superscript $^{\circ}$ to make later expressions less cumbersome). $\boldsymbol{v}_{\rm orbital}$ is the Keplerian velocity vector. The model of \citet{Lynch21} was based on the eccentric shearing box and Equation \ref{magnetic field solution} was setup to satisfy the (approximate) solenoidal condition of the local model (Equation C17 of \citet{Ogilvie14}). One can show that it also satisfies the exact solenoidal condition by adopting the $(a,M,\tilde{z})$ coordinate system and taking the divergence,

\begin{equation}
\nabla_i B^i = \frac{1}{J_3} \frac{\partial}{\partial M} \left ( J_3 B_{t 0} (a,z/H) \frac{J^{\circ}}{J} \frac{H^{\circ}}{H} \right) + \frac{\partial}{\partial \tilde{z}} \left(  B_{z 0} (a) \frac{J^{\circ}}{J} \right) = 0 ,
\end{equation}
as a result of the independence of $J_3 B^M$ (first term) and $B^z$ (second term) on $M$ and $\tilde{z}$ respectively, and we have used $v_{\rm orbital}^{i} = n \hat{e}_M^i$ in this coordinate system.

In cylindrical, unstratified, geometry $H/H^{\circ} = 1$ and $B_{t 0} (a,z/H) = B_{t 0} (a)$. Making use of a dimensionless Jacobian determinant $j = J/J^{\circ}$ we obtain the following for the magnetic pressure due to the magnetic field given by Equation \ref{magnetic field solution},

\begin{equation}
\frac{B^2}{2 \mu_0} = a^2 \frac{B_{t 0}^2 (a)}{2 \mu_0} j^{-2} \frac{1 + e \cos E}{1 - e \cos E} + \frac{B_{z 0}^2}{2 \mu_0} (a) j^{-2} ,
 \label{mag pre}
\end{equation} 
where we have made use of $v_{\rm orbital}^2 = a^2 n^2 \frac{1 + e \cos E}{1 - e \cos E}$.

Substituting Equation \ref{mag pre} for the magnetic pressure into the Lagrangian we arrive at 

\begin{align}
\begin{split}
L &= \iint \Sigma^{\circ} \Biggl[ \frac{1}{2} u^2 - \Phi(\bar{\boldsymbol{x}}) - \bar{\varepsilon} (\bar{\boldsymbol{a}}, J) \\
&- a^2 \frac{1 + e \cos E}{1 - e \cos E} \frac{(J/J^{\circ})^{-1} B^2_{t 0} (\bar{\boldsymbol{a}}) H^{\circ}}{2 \mu_0 \Sigma^{\circ}} - \frac{(J/J^{\circ})^{-1} B^2_{z 0} (\bar{\boldsymbol{a}}) H^{\circ}}{2 \mu_0 \Sigma^{\circ}} \Biggr]  d^2 \bar{\boldsymbol{a}} .
\end{split}
\label{2d mhd lagrangian}
\end{align}
Expanding Equation \ref{2d mhd lagrangian} into the Keplerian Lagrangian $L_{K}$ and a perturbation,
\begin{align}
\begin{split}
L &= L_k - \iint \Sigma^{\circ} \Biggl[ V(\bar{\boldsymbol{x}}) + \bar{\varepsilon} (\bar{\boldsymbol{a}}, J) \\
&+ a^2 \frac{1 + e \cos E}{1 - e \cos E} \frac{(J/J^{\circ})^{-1} B^2_{t 0} (\bar{\boldsymbol{a}}) H^{\circ} }{2 \mu_0 \Sigma^{\circ}} + \frac{(J/J^{\circ})^{-1} B^2_{z 0} (\bar{\boldsymbol{a}}) H^{\circ}}{2 \mu_0 \Sigma^{\circ}} \Biggr]  d^2 \bar{\boldsymbol{a}} ,
\end{split}
\label{lagangian init}
\end{align}
where $V(\boldsymbol{x}) = \Phi (\boldsymbol{x}) - \Phi_K (\boldsymbol{x})$. At leading order we have,

\begin{equation}
\frac{\delta L}{\delta \boldsymbol{x}} \approx \frac{\delta L_{K}}{\delta \boldsymbol{x}} = 0,
\end{equation}
which is Keplerian orbital motion in the plane. Performing the Whitham/orbit average \citep{Whitham65} of Equation \ref{lagangian init}, using the fact that the Lagrangian is nearly integrable,

\begin{align}
\begin{split}
L &= \int m_{a} \left[ n a^2 \dot{M} + n a^2 \sqrt{1 - e^2} \dot{\varpi} + \frac{G M_1}{2 a}\right] \, d a \\
& - \int m_{a} \Biggl[ \langle V(\bar{\boldsymbol{x}}) \rangle + \langle \bar{\varepsilon} (\bar{\boldsymbol{a}}, J) \rangle \\
&+ a^2 \frac{ B^2_{t 0} (\bar{\boldsymbol{a}}) H^{\circ}}{2 \mu_0 \Sigma^{\circ}}  \left \langle \frac{1 + e \cos E}{1 - e \cos E} j^{-1} \right \rangle + \frac{ B^2_{z 0} (\bar{\boldsymbol{a}}) H^{\circ} }{2 \mu_0 \Sigma^{\circ}} \langle j^{-1} \rangle \Biggr]  \, d a ,
\end{split}
\label{orbit avded L}
\end{align}
where $\langle \cdot \rangle = \frac{1}{2 \pi} \int \cdot \, d M$ denotes an orbit average and we have introduced the mass per unit semimajor axis, $m_a$, which is related to the surface density in the reference circular disc by

\begin{equation}
 m_a = 2 \pi a \Sigma^{\circ} .
\end{equation}

The dynamics of $M$ is dominated by orbital motion. Therefore we separate out the terms in the Lagrangian describing the orbital motion from those describing the slow evolution of the disc orbits,

\begin{align}
\begin{split}
L &=  \int m_{a} n a^2 \sqrt{1 - e^2} \dot{\varpi}  \, d a - \int m_{a} \Biggl[ \langle V(\bar{\boldsymbol{x}}) \rangle + \langle \bar{\varepsilon} (\bar{\boldsymbol{a}}, J) \rangle \\
&+ a^2 \frac{ B^2_{t 0} (\bar{\boldsymbol{a}}) H^{\circ}}{2 \mu_0 \Sigma^{\circ}} \left \langle \frac{1 + e \cos E}{1 - e \cos E} j^{-1} \right \rangle + \frac{ B^2_{z 0} (\bar{\boldsymbol{a}}) H^{\circ}}{2 \mu_0 \Sigma^{\circ}} \langle j^{-1} \rangle \Biggr]  \, d a .
\end{split}
\end{align}
The associated Hamiltonian is obtained via a Legendre transform,

\begin{align}
\begin{split}
\mathcal{H} &= \int m_{a} n a^2 \sqrt{1 - e^2} \dot{\varpi}  \, d a - L \\
&= \int m_{a} \Biggl[ \langle V(\bar{\boldsymbol{x}}) \rangle + \langle \bar{\varepsilon} (\bar{\boldsymbol{a}}, J) \rangle \\
&+ a^2 \frac{ B^2_{t 0} (\bar{\boldsymbol{a}}) H^{\circ}}{2 \mu_0 \Sigma^{\circ}} \left \langle \frac{1 + e \cos E}{1 - e \cos E} j^{-1} \right \rangle + \frac{ B^2_{z 0} (\bar{\boldsymbol{a}}) H^{\circ}}{2 \mu_0 \Sigma^{\circ}} \langle j^{-1} \rangle \Biggr] \,  d a .
\end{split}
\end{align}

Henceforth we shall only consider Keplerian potentials so that $V(\bar{\boldsymbol{x}})  = 0$. For a perfect gas we can write \citep{Ogilvie19},

\begin{equation}
m_{a} \langle \bar{\varepsilon} (\bar{\boldsymbol{a}}, J) \rangle = H_{a}^{\circ} F^{(\gamma)} ,
\end{equation}
where we have introduced the geometric part of the Hamiltonian,

\begin{equation}
F^{(p)} = \frac{1}{p - 1} \langle j^{-(p - 1)} \rangle ,
\end{equation}
along with the circular Hamiltonian density,

\begin{equation}
 H_{a}^{\circ} = 2 \pi a P_g^{\circ} ,
\end{equation}
where $P_g^{\circ}$ is the vertically integrated gas pressure in the reference disc. Introducing the Alfv\'{e}n velocity in the reference disc: $v^{i}_{a 0} = B^{i}_0/\sqrt{\mu_0 \rho^{\circ}}$, we parameterise the magnetic field strength in terms of a dimensionless toroidal $(V_t)$ and vertical $(V_z)$ Alfv\'{e}n velocities, where $V_t = a v^{E}_{a 0}/c_s$ and $V_z =  v^{z}_{a 0}/c_s$. 

From \citet{Ogilvie19} we have the following expression for $F^{(2)}$,

\begin{equation}
 F^{(2)} (e, q, E_0) = \langle j^{-1} \rangle = \frac{\sqrt{1 - e^2}}{1 - e (e + a e_a)} \frac{q - e (1 - \sqrt{1 - q^2}) \cos E_0}{q \sqrt{1 - q^2}} .
\end{equation}
Similarly we obtain can obtain an expression for $\left \langle \frac{1 + e \cos E}{1 - e \cos E} j^{-1} \right \rangle$,

\begin{align}
\begin{split}
 \left \langle \frac{1 + e \cos E}{1 - e \cos E} j^{-1} \right \rangle &= \frac{\sqrt{1 - e^2}}{1 - e (e + a e_a)} \frac{q + e (1 - \sqrt{1 - q^2}) \cos E_0}{q \sqrt{1 - q^2}} \\
&=  F^{(2)} (e, q, E_0 + \pi) .
\end{split}
\end{align}
Thus the vertical magnetic field acts like a $\gamma=2$ perfect gas, while the quasi-toroidal magnetic field acts like a $\gamma=2$ gas with an anti-phased orbital compression.

We can then write the Hamiltonian as

\begin{align}
\begin{split}
\mathcal{H} = \int H_{a}^{\circ} \Biggl( F^{(\gamma)} (e,q,E_0) &+ \frac{\gamma}{2} V_t^2 F^{(2)} (e,q,E_0 + \pi) \\
&+ \frac{\gamma}{2} V_z^2 F^{(2)} (e,q,E_0)  \Biggr)  \, d a , \label{seperated hamiltonian}
\end{split}
\end{align}
where the factor of $\gamma$ in the magnetic terms appear as a result of factoring out $H^{\circ}_a$. This Hamiltonian can be split into contributions from the gas internal energy, and the energy in the quasi-toroidal and vertical magnetic fields, with $\mathcal{H} = \mathcal{H}_{\rm gas} + \mathcal{H}_{\rm tor} + \mathcal{H}_{\rm vert}$, corresponding to the first second and third term in the brackets of Equation \ref{seperated hamiltonian}. As in gas discs each of these terms are a product of the, geometry independent, Hamiltonian density in the reference disc and a geometric part which encapsulates the dependence on the orbital geometry.

It is convenient to reformulate this Hamiltonian as a Hamiltonian for a single effective gas. We can do this by introducing a new geometric part of the Hamiltonian,

\begin{align}
\begin{split}
&F_{V_t, V_z}  (e,q,E_0) = \frac{1}{1 + \frac{\gamma}{2} (V_t^2 + V_z^2)} F^{(\gamma)} (e,q,E_0)  \\
&+ \frac{\frac{\gamma}{2} V_t^2}{1 +  \frac{\gamma}{2} (V_t^2 + V_z^2)} F^{(2)} (e,q,E_0 + \pi)   + \frac{\frac{\gamma}{2} V_z^2}{1 +  \frac{\gamma}{2} (V_t^2 + V_z^2)} F^{(2)} (e,q,E_0)  ,
\end{split}
\end{align}
which is a weighted sum of two adiabatic gas $F^{(p)}$ with different ratio of specific heats and a third nonadiabatic gas term for the toroidal field. In this case the Hamiltonian can be written as,

\begin{equation}
\mathcal{H} = \int \tilde{H}_{a}^{\circ} F_{V_t, V_z} (e,q,E_0) d a,
\end{equation}
where we have introduced

\begin{equation}
\tilde{H}_{a}^{\circ} = 2 \pi a (P^{\circ}_g + P_m^{\circ}) = \left(1 + \frac{\gamma}{2} V_t^2 + \frac{\gamma}{2} V_z^2 \right) H_{a}^{\circ} = \frac{1 + \beta^{\circ}}{\beta^{\circ}} H_a^{\circ},
\end{equation}
which is the Hamiltonian density in the reference circular disc and $\beta^{\circ}$, the plasma-$\beta$ in the reference circular disc. Unlike the simpler perfect gas case $F_{V_t, V_z}$ is no longer only a function of the geometry for a given ratio of specific heat, it also depends on the ``partial pressure'' of the constitutive effective gasses on a given orbit.

Hamilton's equations in the noncannoical $e(a,t)$, $\varpi(a,t)$ are \citep{Ogilvie19}

\begin{align}
 m_a \frac{\partial e}{\partial t} &= \frac{\sqrt{1 - e^2}}{n a^2 e} \frac{\delta \mathcal{H}}{\delta \varpi} , \label{d e} \\
  m_a \frac{\partial \varpi}{\partial t} &= -\frac{\sqrt{1 - e^2}}{n a^2 e} \frac{\delta \mathcal{H}}{\delta e} . \label{d varpi}
\end{align}

The ideal MHD eccentric disc Hamiltonian preserves the symmetries of the unmagnetised eccentric disc Hamiltonian of \citet{Ogilvie19}; i.e. time translation and global rotation, with the Hamiltonian only depending on $\varpi$ through it derivative $\varpi_{a}$. As such, following \citet{Ogilvie19}, one can show that the total Hamiltonian, $\mathcal{H}$, which in an unstratified disc corresponds to the sum of the magnetic and internal energies, and the angular momentum deficit (AMD), a positive definite measure of the total eccentricity commonly used in celestial mechanics,

\begin{equation}\label{eq:AMD}
 \mathcal{C} = \int m_{a} n a^2 \left(1 - \sqrt{1 - e^2} \right) d a ,
\end{equation}
are conserved.

The simplest solutions to the eccentric disc equations are the eccentric modes, which are solutions where $e$ is independent of time and the disc is untwisted and uniformly precessing at an angular frequency $\omega$. These solutions are a particularly convenient setting for numerical simulations as they avoid strong differential precession seen in generic eccentricity profiles. The eccentric mode equation is obtained from Equation \ref{d varpi},

\begin{equation}
   m_a \omega = -\frac{\sqrt{1 - e^2}}{n a^2 e} \frac{\delta \mathcal{H}}{\delta e} . \label{mode eq}
\end{equation}
Equation \ref{d e} is automatically satisfied as the disc is untwisted (so $\frac{\delta \mathcal{H}}{\delta \varpi} = 0$) and $e$ is independent of time.

As shown in \citet{Ogilvie19}, Equation \ref{mode eq} can be written as

\begin{equation}
 \omega \frac{\delta L}{\delta e} = \frac{\delta \mathcal{H}}{\delta e} , \label{variational form}
\end{equation}
where 

\begin{equation}
    L = \int m_a n a^2 \sqrt{1 - e^2} \, d a 
\end{equation}
is the angular momentum. Equation \ref{variational form} can be interpreted as a variational problem which makes $\mathcal{H}$ (here corresponding to the total disc internal + magnetic energy) stationary at a fixed angular momentum $L = \mathrm{const}$.

For constant $V_t, V_z$ the eccentric mode equations are explicitly \citep{Ogilvie19}:

\begin{align}
\begin{split}
- \frac{\omega m_a}{\tilde{H}_{a}^{\circ}} \frac{n a^2 e}{\sqrt{1 - e^2}} &= \frac{\partial F_{V_t, V_z}}{\partial e} - a e_a \frac{\partial^2 F_{V_t, V_z}}{\partial e \partial f} \\
&- a (2 e_a + a e_{a a}) \frac{\partial^2 F_{V_t, V_z}}{\partial f^2} - \frac{d \ln (\tilde{H}^{\circ}_a)}{d \ln a} \frac{\partial F_{V_t, V_z}}{\partial f} ,
\end{split}
\label{constant beta eccentric mode}
\end{align}
where we have introduced $f = e + a e_a$. If $V_t$ or $V_z$ depends on the semimajor axis then the equation for an eccentric mode becomes,

\begin{align}
\begin{split}
- \frac{\omega m_a}{\tilde{H}_{a}^{\circ}} \frac{n a^2 e}{\sqrt{1 - e^2}} &= \frac{\partial F_{V_t, V_z}}{\partial e} - a e_a \frac{\partial^2 F_{V_t, V_z}}{\partial e \partial f} \\
&- a (2 e_a + a e_{a a}) \frac{\partial^2 F_{V_t, V_z}}{\partial f^2} - \frac{d \ln ( H^{\circ}_a)}{d \ln a} \frac{\partial F_{V_t, V_z}}{\partial f} \\
&- \frac{\gamma}{2} \frac{\beta^{\circ}}{1 + \beta^{\circ}} \left( \frac{d V_t^2 }{d \ln a} \frac{\partial F^{(2)}}{\partial f} \Biggl |_{E_0=\pi} + \, \frac{d V_z^2 }{d \ln a} \frac{\partial F^{(2)}}{\partial f} \Biggl |_{E_0=0} \right) .
\end{split} \label{variable beta eccentric mode}
\end{align}
Note that we have the perfect gas circular Hamiltonian density ($H_{a}^{\circ}$) on the forth term on the right hand side. We also have $F^{(2)} |_{E_0 = 0} = F^{(2)} (e,q(e,f), 0)$ and $F^{(2)} |_{E_0 = \pi} = F^{(2)} (e,q(e,f), \pi)$.

For untwisted discs $F_{V_t, V_z} (e, f)$ has an apparent singularity when $e=f$, where the eccentricity gradients vanish. Following \citet{Ogilvie19} this apparent singularity can be removed using the trigonometric parametrisation $e = \sin 2 \alpha$, $f = \sin 2 \beta$. Expressions for $F^{(1)}(e,q,0)$ and $F^{(2)} (e,q,0)$, in terms of this parametrisation are given in Appendix C of \citet{Ogilvie19}. For including quasi-toroidal fields we will also need 

\begin{equation}
 F^{(2)} (e,q,\pi) = \cos(2 \alpha) \cos(\alpha - \beta) \sec(2 \beta) \sec(\alpha + \beta) .
\end{equation}

\section{Linear Theory} \label{linear theory}

When $e$, $a e_a$ and $a e \varpi_a$ are much less than unity, the geometric part of the Hamiltonian density in a 2D disc can be approximated as \citep{Ogilvie19}

\begin{equation}
F^{(\rm 2D)} \approx \frac{1}{2} e (e + a e_a) + \frac{1}{4} \gamma \left[ (a e_a)^2 + (a e \varpi_a)^2 \right] , \label{linear F}
\end{equation}
where we have dropped an unimportant constant term, that has no influence on the dynamics, so that we can use this expression for isothermal discs. In addition to Equation \ref{linear F} we require the linear limit of $F^{(2)} (e,q,E_0 + \pi)$, to include the quasi-toroidal field, this can be obtained in a similar way and is

\begin{equation}
F^{(2)} (e,q,E_0 + \pi) \approx \frac{1}{2} e (e + a e_a) + \frac{1}{2} \left[ (a e_a)^2  + (a e \omega_a)^2\right] + a e e_a ,
\end{equation}
where the first two terms arise from the adiabatic variation of the magnetic pressure, similar to the vertical field, while the last term arises from the magnetic tension and the non-adiabatic variation of the magnetic pressure. Combining these we arrive at an expression for $F_{V_t, V_z}$,

\begin{align}
\begin{split}
F_{V_t, V_z}  (e,q,E_0) &\approx \frac{1}{2} e (e + a e_a) + \frac{\tilde{\gamma}}{4} \left[ (a e_a)^2  + (a e \omega_a)^2\right]  \\
&+ \frac{\frac{\gamma}{2} V_t^2}{1 +  \frac{\gamma}{2} (V_t^2 + V_z^2)} a e e_a ,
\end{split} \label{combined linear theory}
\end{align}
where we have introduced a modified ratio of specific heats

\begin{equation}
\tilde{\gamma} = \gamma \frac{1 +  V_t^2 + V_z^2}{1 +  \frac{\gamma}{2} (V_t^2 + V_z^2)}  .
\end{equation}
To connect with the existing work on linear eccentric disc theory it is useful to rewrite Equation \ref{combined linear theory} in terms of the complex eccentricity $\mathcal{E} = e \exp(i \varpi)$, the geometric part of the Hamiltonian in the linear limit is then

\begin{align}
\begin{split}
F_{V_t, V_z}  (e,q,E_0) &\approx \frac{1}{2} [|\mathcal{E}|^2 + \mathrm{Re}(a \mathcal{E} \mathcal{E}^{*}_a)] + \frac{\tilde{\gamma}}{4} |a \mathcal{E}_a |^2 \\
&+ \frac{\frac{\gamma}{2} V_t^2}{1 +  \frac{\gamma}{2} (V_t^2 + V_z^2)} \mathrm{Re}(a \mathcal{E} \mathcal{E}^{*}_a) .
\end{split}
\end{align}
The, non-canonical, Hamilton's equations for the complex eccentricity are\footnote{This form of Hamilton's equations for the eccentric disc theory was originally suggested by Prof. Gordon Ogilvie in an earlier draft of \citet{Ogilvie19} as a way of connecting Hamiltonian eccentric disc theory with the Shr\"{o}dinger equation.}

\begin{equation}
m_a \dot{\mathcal{E}} = -\frac{2 i \sqrt{1 - e^2}}{n a^2} \frac{\delta \mathcal{H}}{\delta \mathcal{E}^{*}} , \label{Complex E Hamiltonian}
\end{equation}
with the functional derivative of $\mathcal{E}^{*}$ being related to the functional derivative of $e$ and $\varpi$ through,

\begin{equation}
 \frac{\delta}{\delta \mathcal{E}^{*}} = \frac{1}{2} \mathrm{e}^{i \varpi} \left(\frac{\delta}{\delta e} + \frac{i}{e} \frac{\delta}{\delta \varpi}  \right) .
\end{equation}
Substituting in the Linear form of the Hamiltonian into Equation \ref{Complex E Hamiltonian} and performing the functional derivative we obtain a linear equation for the evolution of the complex eccentricity in an unstratified ideal MHD disc,

\begin{align}
\begin{split}
2 \Sigma^{\circ} n a^3 \frac{\partial \mathcal{E}}{\partial t} &= \frac{\partial}{\partial a} \left( i \tilde{\gamma} P^{\circ} a^3 \frac{\partial \mathcal{E}}{\partial a}  \right) + i a^2 \frac{d P^{\circ}}{d a} \mathcal{E} + i \partial_a \left[ a^2 \frac{(a B_0^{\phi})^2}{\mu_0} \right]  \mathcal{E}
\end{split} 
\end{align}
where we have used $M_a = 2 \pi a \Sigma^{\circ}$ and $\tilde{H}_a^{\circ} = 2 \pi a P^{\circ}$. The first two terms on the right-hand side correspond to an adiabatic gas with an effective ratio of specific heats, $\tilde{\gamma}$ set by the plasma-$\beta$. The final term arises from the non-adiabatic change to the magnetic pressure from the stretching of the magnetic field lines\footnote{This can be shown by deriving the linear, magnetised, eccentric disc equations following a similar procedure to \citet{Goodchild06}, a task that is significantly more involved than taking the linear limit of the Hamiltonian theory.}. Terms arising due to the magnetic tension cancel with additional non-adiabatic terms along with the modification to the background rotation profile as a result of the magnetic tension in the unperturbed disc.  

Specialising to an eccentric mode in a disc with a purely vertical field ($V_t = 0$) the equation simplifies to

\begin{equation}
2 \Sigma^{\circ} n a^3 \omega \mathcal{E} = \frac{\partial}{\partial a} \left( \tilde{\gamma} P^{\circ} a^3 \frac{\partial \mathcal{E}}{\partial a}  \right)  
+  \frac{d  P^{\circ}}{d a} a^2 \mathcal{E} .
\end{equation}
When the gas pressure, $P_g$, and dimensionless Alfv\'{e}n velocity, $V_z$, are constant one can rewrite the above equation as

\begin{equation}
2 \Sigma^{\circ} n a^3 \tilde{\omega} \mathcal{E} = \frac{\partial}{\partial a} \left( \gamma P_g^{\circ} a^3 \frac{\partial \mathcal{E}}{\partial a}  \right)  ,
\end{equation}
where we have introduced a rescaled precession frequency $\tilde{\omega} = \omega/(1 + V_z^2)$. Therefore, under these restrictions on the pressure and magnetic field, the eccentric mode in the magnetised and unmagnetised discs are identical and differ only by their precession frequency.

\section{Nonlinear Modes in an Isothermal Disc} \label{nonlinear modes}

Our primary motivation for extending the unstratified eccentric disc theory to include magnetic fields is to provide initial conditions for simulations of the eccentric MRI. As such we focus on calculating the nonlinear eccentric modes found in a disc contained between two, circular, rigid walls as done to setup the hydrodynamical simulations of \citet{Barker16}, rather than the more realistic free boundaries considered in \citet{Ogilvie19}. As a model of a realistic MHD disc, however, the unstratified model derived in the previous section has major limitations; namely it fails to account for the dynamical vertical structure of the disc which is known to be important to correctly describe the dynamics of eccentric discs \citet{Ogilvie01,Ogilvie08,Ogilvie14,Ogilvie19} and can significantly increase the field strength of the quasi-toroidal field, particularly for more nonlinear eccentric discs \citep{Lynch21}. There is also the issue of how the disc interacts with the external magnetic fields.

We, thus, consider a simple MHD generalisation of the mode computed in \citet{Barker16}. This consists of a globally isothermal disc with a constant reference surface density, $\Sigma^{\circ}$, with the Hamiltonian density in the reference circular disc being $H^{\circ}_a = 2 \pi a c_s^2 \Sigma^{\circ}$, where $c_s$ is a constant sound speed.

For the modes that will be simulated in Section \ref{simulations} we impose a purely vertical field with,
\begin{equation}
 V_t = 0 , \quad V_z = \frac{n l_{\rm MRI}}{2 \pi c_s \sqrt{16/15}} W(a) .
 \label{vz case}
\end{equation}
For comparison, in this section, we also compute a quasi-toroidal case with

\begin{equation}
V_t = \frac{n l_{\rm MRI}}{2 \pi c_s \sqrt{16/15}} W(a) , \quad V_z = 0 .
\label{vt case}
\end{equation}
In both cases $l_{\rm MRI}$ is a constant lengthscale, which in the vertical field case corresponds to the lengthscale of the fastest growing MRI mode in the reference circular disc. $W(a)$ describes the taper on the inner and outer disc boundaries, for which we use

\begin{equation}\label{eq:env}
W (a) = \frac{1}{2} \left[1 + \tanh \left(\frac{a - a_{\rm in}}{w_{\rm transition}} \right) \right] \left[1 - \tanh \left(\frac{a - a_{\rm out}}{w_{\rm transition}} \right) \right] ,
\end{equation}
when we wish to include a taper. The disc is contained within two rigid circular walls located at $a_{\rm min}$ and $a_{\rm max}$, such that $a_{\rm min} \le a_{\rm in} \le a_{\rm out} \le a_{\rm max}$. We therefore have boundary conditions $e (a_{\rm min}) = e (a_{\rm max}) = 0$. The precession frequency of the mode, $\omega$, is an eigenvalue of the problem. One can solve Equation \ref{variable beta eccentric mode}for the eccentric mode by specifying $e_a$ on the inner boundary and employing a shooting method to obtain $\omega$.

\begin{figure}
\includegraphics[width=\linewidth]{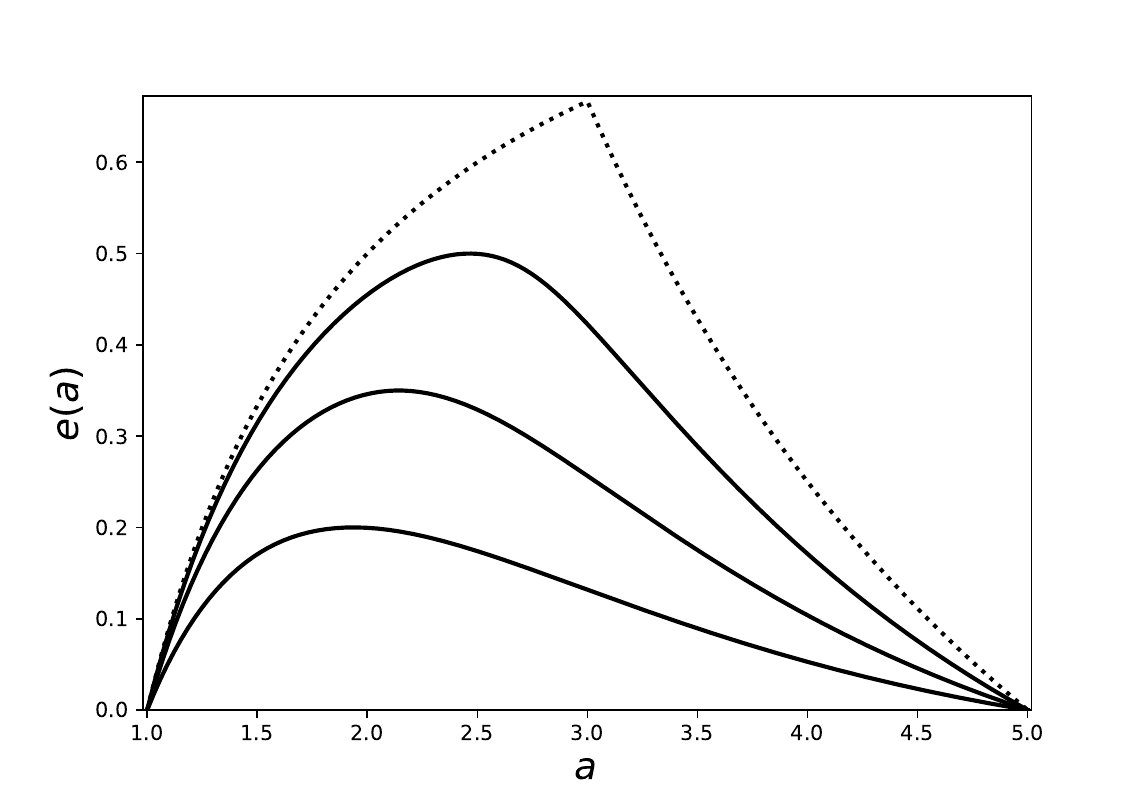}
\caption{Eccentricity profiles of the eccentric modes used in simulations with $\max(e) = 0.2$, $\max(e) = 0.35$ and $\max(e) = 0.5$ respectively. For both modes $l_{\rm MRI} = c_s/n(a_{\rm min})$, $a_{\rm min} = 1$, $a_{\rm in} = 1.5$ , $a_{\rm out} = 4.5$, $a_{\rm max} = 5$ and $w_{\rm transition} = 0.05$. The eccentric modes in the unmagnetised and MHD discs with a purely vertical field are almost indistinguishable. The dotted line is the ``limiting slope" solution.}
\label{mode for sim}
\end{figure}

To provide initial conditions for our simulations we solve for eccentric modes with $l_{\rm MRI} = c_s/n(a_{\rm min})$, $a_{\rm min} = 1$, $a_{\rm in} = 1.5$, $a_{\rm out} = 4.5$, $a_{\rm max} = 5$ and $c_s = 0.05$, with purely vertical fields. These modes are shown in Figure \ref{mode for sim}. The simulated modes with different choices of parameters (as discussed in Section \ref{simulations}) have functionally indistinguishable eccentricity profiles.

The MHD eccentric modes depicted in Figure \ref{mode for sim} are nearly indistinguishable from the unmagnetised case. The same is true of eccentric modes computed using a similar strength quasi-toroidal field. This is perhaps not surprising given the magnetic field strength in these discs is set by the requirement that the circular reference disc is MRI unstable. This means the plasma-$\beta$ in the reference disc never drops below $\beta=84$. In the eccentric disc the range of plasma-$\beta$ attained is greater as a result of the lateral orbital compression which occurs in the presence of eccentricity gradients. 

\begin{figure}
\includegraphics[width=\linewidth]{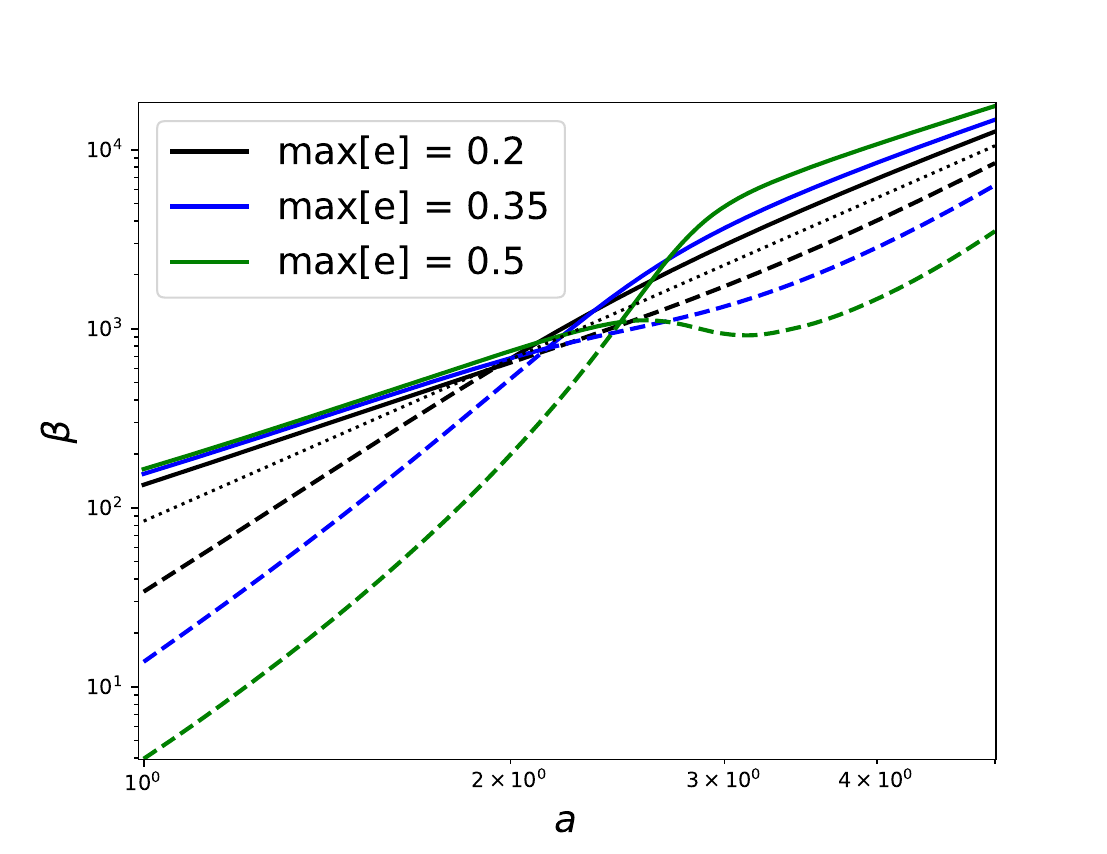}
\caption{Minimum (dashed lines) and maximum (solid lines) plasma beta on orbits for the eccentric modes with a purely vertical field and no taper. The dotted line is the plasma beta in the circular disc. Orbital compression can lead to significant local enhancement of the magnetic fields, while at the eccentricity maxima the plasma-$\beta$ is constant around the orbit. The simulations do not reach as extreme plasma-$\beta$s as a result of the taper.}
\label{plasma beta bz}
\end{figure}

\begin{figure}
\includegraphics[width=\linewidth]{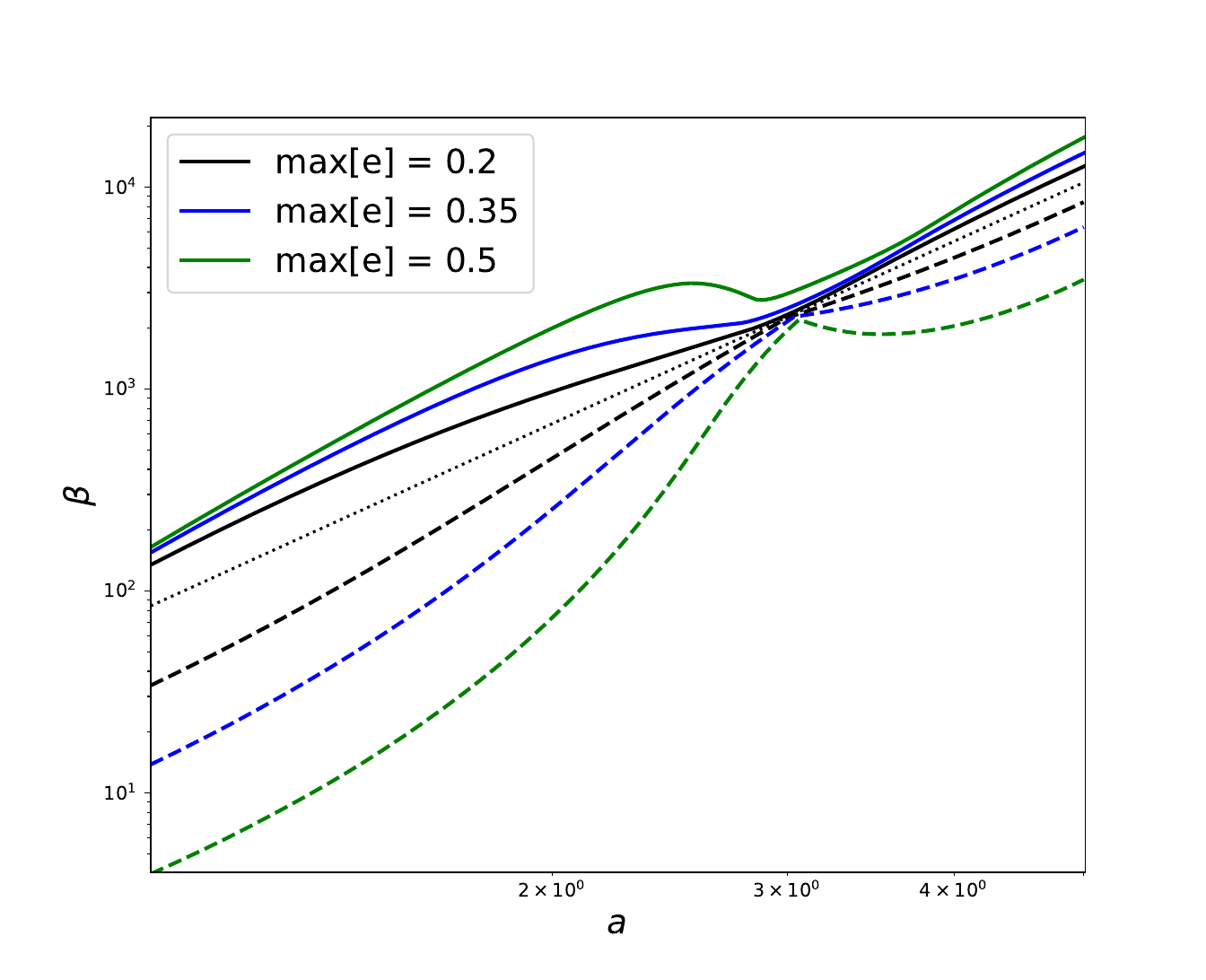}
\caption{Same as Figure \ref{plasma beta bz} but for a quasi-toroidal field. There is a greater variation of the magnetic field around the orbit in the inner disc. Unlike the vertical field case the stretching of the field lines around an eccentric orbit means the magnetic field always varies around the orbit, even at the eccentricity maxima where the surface density is constant. }
\label{plasma beta bt}
\end{figure}

Figures \ref{plasma beta bz} and \ref{plasma beta bt} show the minimum and maximum plasma-$\beta$ on an orbit, in the absence of a taper, for the vertical and quasi-toroidal field models respectively. In the absence of a taper the orbital compression results in regions of high magnetic field strengths in the inner disc, particularly for the mode with $\max[e]=0.5$ . This effect is lessened when the taper is included as the magnetic field strength drops to zero close to the boundary where the effects of orbital compression are greatest. Despite attaining plasma-$\beta$ as low as $\beta \sim 4$ for the $\max[e] = 0.5$ mode, without taper, differs only slightly from the unmagnetised case (having a slightly lower eccentricity gradient on the inner boundary). This is, in part, a geometric effect where the shape of highly nonlinear eccentric mode is dictated by the requirement that $|q|<1$ to avoid an orbital intersection. As discussed in \citet{Barker16,Ogilvie19} this results in a limiting slope solution given by

\begin{equation}
a e = \begin{cases}
         a - a_{\rm min} , & a_{\rm min} < a < \bar{a} \\
         -a + a_{\rm max} , & \bar{a} < a < a_{\rm max}
         \end{cases}
\end{equation}
where $\bar{a} = (a_{\rm min} + a_{\rm max})/2$. In principle higher order limiting slope solutions might depend on the magnetic field strength as it is not obvious how the mode selects nodes for the higher order modes.

The large orbital compression responsible for the regions of greatly enhanced magnetic fields in the modes calculated above are primarily a consequence of the adoption of rigid circular boundaries. As shown in \citet{Ogilvie19} adoption of more realistic free boundaries conditions result in more moderate eccentricity gradients for a given $\max[e]$. In Appendix \ref{free boundaries} we solve for the MHD eccentric modes with free boundaries and a taper in both the magnetic field and the surface density. The variation of the magnetic field strength and the orbit are much reduced compared with the equivalent rigid boundary eccentric mode due to the smaller eccentricity gradients. This confirms that the strong enhancement of the magnetic fields seen in the modes computed for Figures \ref{plasma beta bz} and \ref{plasma beta bt} are primarily a consequence of the rigid wall boundaries.

\begin{figure}
\includegraphics[width=\linewidth]{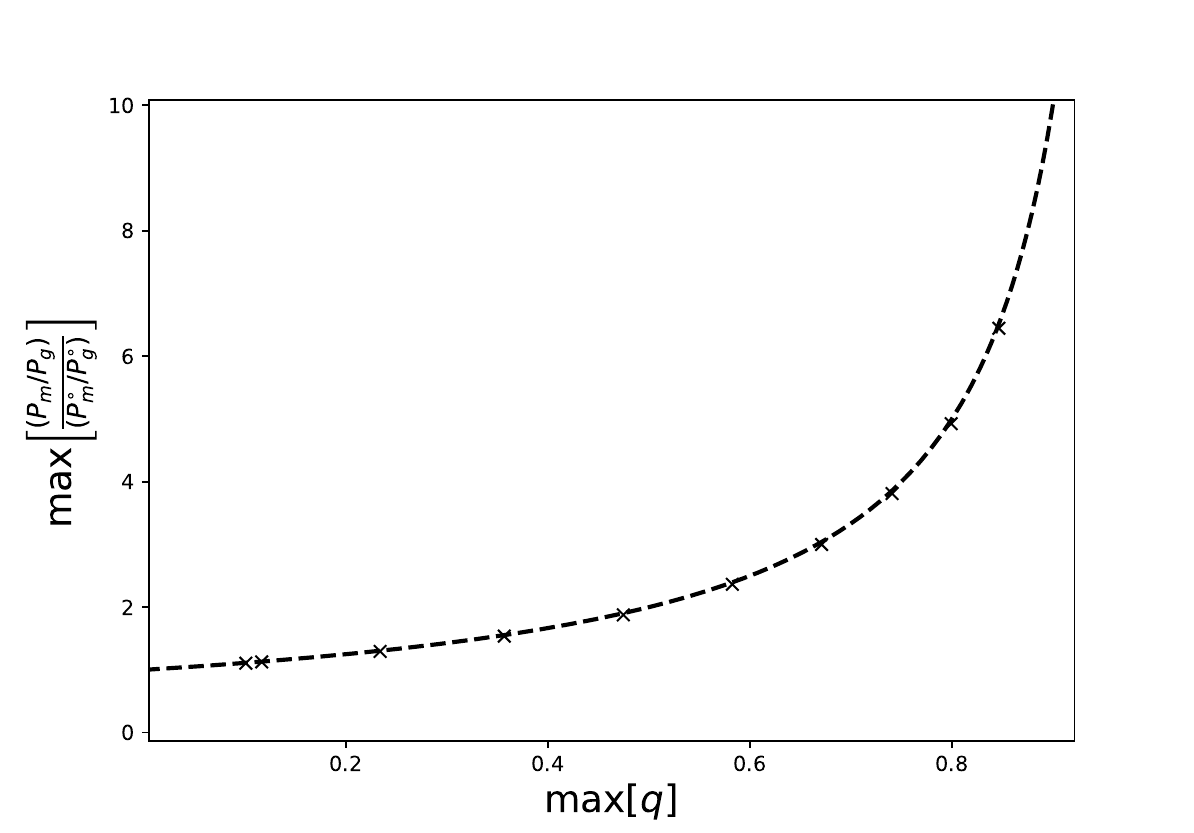}
\caption{Mode number dependence of the magnetic field enhancement for modes with $\max[e]=0.1$. Dashed line shows the function $1/(1 - q_{\rm max})$, which is the approximate magnetic field enhancement for disc with low eccentricity. }
\label{plasma beta mode dependance}
\end{figure}

While the strong enhancement of the magnetic field in the fundamental (zero node) modes considered thus far are primarily a consequence of the choice of boundary conditions, higher order modes (i.e. with multiple nodes) will attain larger eccentricity gradients for a given $\max[e]$. Figure \ref{plasma beta mode dependance} shows the maximum magnetic field enhancement for eccentric modes with $\max[e]=0.1$. These modes are discrete due to the boundary conditions. Increasing mode number results in an increased $\max[q]$, resulting in an increasing magnetic field enhancement due to the greater lateral compression. This magnetic field enhancement approximately follows $1/(1 - q_{\rm max})$ for small $\max[e]$. This magnetic field enhancement may have important consequence for the eccentric MRI if the higher field strengths are able to stabilise the MRI.

\section{Nonlinear simulations}\label{simulations}
We have run nonlinear MHD simulations to demonstrate the integrity of the eccentric disc solutions calculated in Section \ref{nonlinear modes}, specifically the vertical field case including a taper. We limit our focus to purely 2D simulations in this paper in order to isolate the eccentric modes' role as MHD equilibria, which are of course unstable in three dimensions. In a companion paper, we explore the growth and turbulent saturation of the instabilities of these equilibria in fully 3D simulations. 

\begin{table*}
\caption{Details of the nonlinear simulations considered in this paper. All simulations were run on a uniform cylindrical mesh with $r_1/r_0=5$, and $c_s=0.05a_\text{min}n(a_\text{min}).$ The magnetized runs include purely vertical fields as described by \eqref{vz case}, with $l_\text{MRI}=c_s/n(a_\text{min})$, $w_\text{transition}=0.1a_\text{min},$ $a_\text{in}=2a_\text{min}$, and $a_\text{out}=4a_\text{min}$ (except for m35t, which uses  $w_\text{transition}=0.05a_\text{min}$, $a_\text{in}=1.5a_\text{min}$, and $a_\text{out}=4.5a_\text{min}$). The final three columns compare precession frequencies predicted by the eccentric mode calculations with frequencies and decay rates measured in the simulations.}
\label{tab:sims}
    \begin{tabular}{cccccccc}
        \hline
        Simulation & $\max[e]$ & $B_z$? & $N_R\times N_\phi$ & 
        Predicted $\omega$ & Simulation $Re[\omega]$ & Simulation $Im[\omega]$ \\
        \hline
        h2   & $0.20$ & No  & $480\times800$ & -0.004235 & -0.0044 & $-4.8\times10^{-5}$ \\
        h35  & $0.35$ & No  & $480\times800$ & -0.005012 & -0.0054 & $-1.0\times10^{-4}$ \\
        h5   & $0.50$ & No  & $480\times800$ & -0.007459 & -0.0045 & $-5.5\times10^{-3}$ \\
        m2   & $0.20$ & Yes & $480\times800$ & -0.004244 & -0.0044 & $-4.8\times10^{-5}$ \\
        m35  & $0.35$ & Yes & $480\times800$ & -0.005042 & -0.0054 & $-1.0\times10^{-4}$ \\
        m35t & $0.35$ & Yes & $480\times800$ & -0.005047 & -0.0055 & $-9.9\times10^{-5}$ \\
        m35l & $0.35$ & Yes & $240\times400$ & -0.005042 & -0.0033 & $-3.8\times10^{-4}$ \\
        m35h & $0.35$ & Yes & $960\times800$ & -0.005042 & -0.0053 & $-6.2\times10^{-5}$ \\
        m5   & $0.50$ & Yes & $480\times800$ & -0.007582 & -0.0045 & $-5.5\times10^{-3}$ \\
        m5h  & $0.50$ & Yes & $960\times800$ & -0.007582 & -0.0046 & $-5.6\times10^{-3}$ 
    \end{tabular}
\end{table*}

\subsection{Setup}
We use a uniform grid version of the code RAMSES \citep{Teyssier2002,Fromang2006,Faure2014}
\footnote{Available at \hyperlink{https://sourcesup.renater.fr/projects/dumses/}{https://sourcesup.renater.fr/projects/dumses/}
}, which employs a high-order Godunov method to solve the magnetohydrodynamic equations under the cylindrical approximation (i.e., without vertical gravity). Taken with a purely isothermal equation of state, these are
\begin{equation}
\label{eq:RAM1}
    \frac{\partial \rho }{\partial t}
    +\nabla\cdot(\rho {\boldsymbol{u}})= 0,
\end{equation}
\begin{equation}\label{eq:RAM2}
    \frac{\partial (\rho {\boldsymbol{u}})}{\partial t}
    +\nabla\cdot( 
        \rho {\boldsymbol{u} \boldsymbol{u}}
        -{\boldsymbol{B} \boldsymbol{B}}
    )
    +\nabla \left(
        P+\frac{\boldsymbol{B}\cdot \boldsymbol{B}}{2}
    \right)
    =-\rho\nabla\Phi,
\end{equation}
\begin{equation}\label{eq:RAM3}
    \frac{\partial {\boldsymbol{B}}}{\partial t}
    +\nabla\cdot({\boldsymbol{u} \boldsymbol{B}}-{\boldsymbol{B} \boldsymbol{u}})
    =0,
\end{equation}
where $\rho$ is the gas density, $P=c_s^2\rho$ is the pressure, $c_s$ is the sound speed, ${\bf u}$ is the fluid velocity field, $\Phi=-GM_1/R$ is the (Newtonian and cylindrical) gravitational potential of a central mass $M$, and ${\bf B}$ is the magnetic field. We initialise purely two-dimensional simulations with the surface densities, radial and azimuthal velocities, and purely vertical magnetic fields (from Equations \ref{eq:rho_from_e}-\ref{eq:Bz_from_e}) corresponding to eccentricity profiles like those shown in Fig. \ref{mode for sim}. Table \ref{tab:sims} lists relevant properties. We do not simulate the quasi-toroidal field case in this paper. This case is complicated by the difficulty in ensuring the solenoidal condition is satisfied when switching from the orbital to the polar grid, this is best done by use of a vector potential (derived in Appendix \ref{appendix vector pontential}) which is not implemented in the version of RAMSES we are using.

We impose quasi-rigid wall boundary conditions at both the inner and outer radial boundaries $r_0=a_\text{min}$ and $r_1=a_\text{max}$, fixing $u_R=0$ and setting $u_\phi$ by the Keplerian angular velocity of the circular reference disk. Nonzero eccentricity gradients at the boundaries imply nonzero surface density variations with $\phi$. We therefore use a zero-gradient boundary condition for the density, setting its value in the ghost cells to the value of the last cell in the active domain. We lastly set $B_z$ to zero in the ghost cells ($B_R$ and $B_\phi$ remain identically zero throughout, and so the magnetic field remains trivially solenoidal in these 2D simulations).

To track the evolution of eccentricity in our simulations, we compute the semimajor axis of each grid cell from \citep[e.g.,][]{Miranda17}
\begin{equation}
    a(R,\phi)=\left(\frac{2}{R} - \frac{u^2}{GM_1}\right)^{-1},
\end{equation}
and the eccentricity vector from
\begin{equation}
    {\boldsymbol{e}}=[e\cos\varpi,e\sin\varpi]
    =(GM_1)^{-1}\left[ 
        u^2{\boldsymbol{R}} - ({\boldsymbol{u}\cdot R}){\boldsymbol{u}}
    \right] 
    -\hat{\boldsymbol{R}}. \label{eq:e from sim}
\end{equation}
Binning the eccentricity values in every cell by semimajor axis, we average within each bin to produce one-dimensional eccentricity profiles $\tilde{e}=\tilde{e}(a)$ at each timestep. These we use in turn to compute the integrated angular momentum deficit $ \mathcal{C}$. We additionally consider the time-evolution of the total Hamiltonian
\begin{equation}\label{eq:H}
\mathcal{H} = \int_{a_{\text{min}}}^{a_{\text{max}}} \! \int_0^{2 \pi} (c_s^2 \Sigma \ln \Sigma + P_m ) \, R \, d \phi\, d R ,
\end{equation}
which is conserved in the ideal secular theory.

\subsection{Results}
\begin{figure*}
\includegraphics[width=\linewidth]{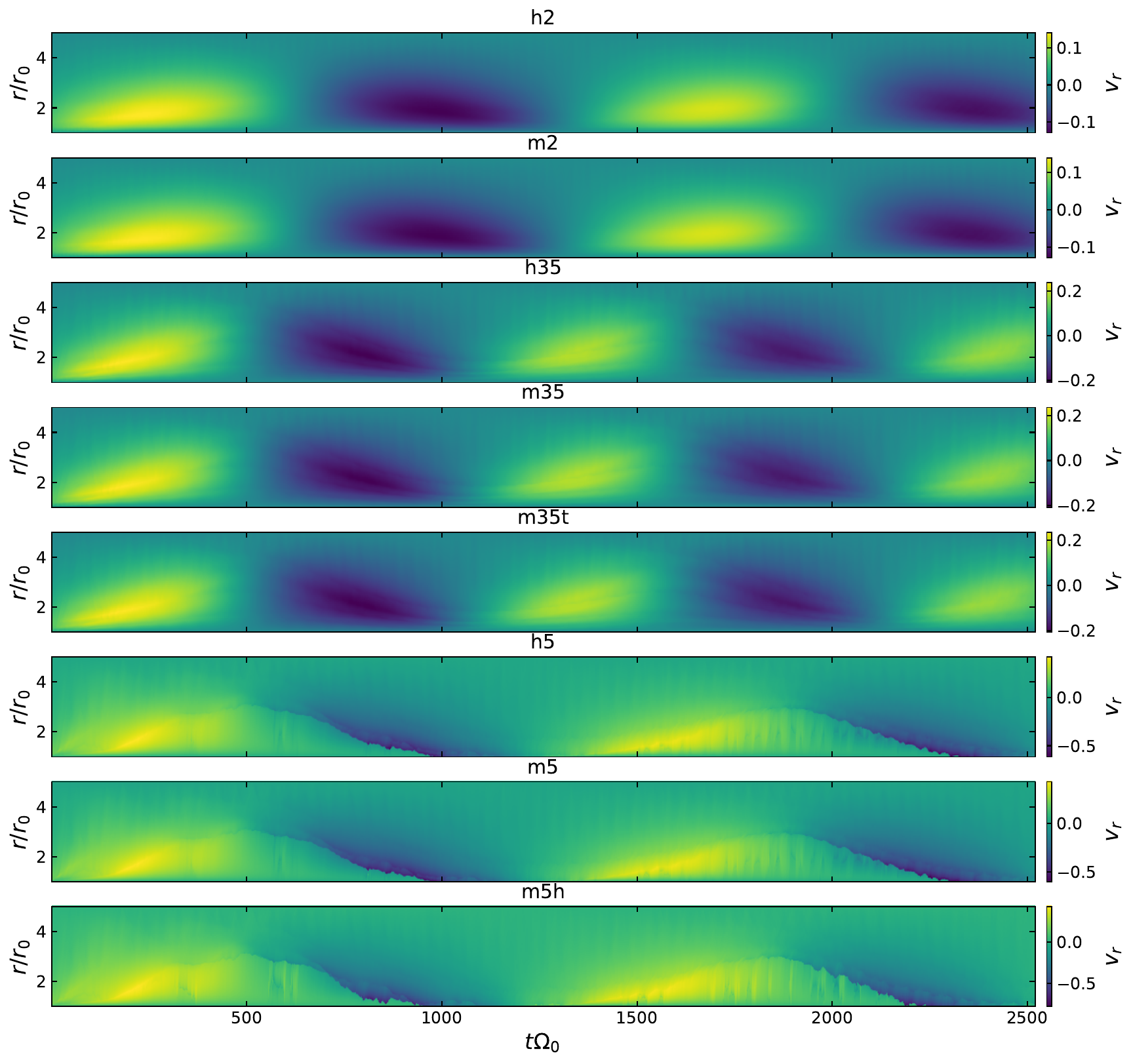}
\caption{Spacetime diagrams showing radial profiles of radial velocity at a fixed $\phi$ vs. time for both hydrodynamic (names starting with $h$) and MHD ($m$) simulations. The periodic changes in the sign of $v_r$ over $t\Omega_0\sim 1000$ illustrate the retrograde precession of the eccentric distortions.}
\label{fig:spcvr}
\end{figure*}

\begin{figure*}
\includegraphics[width=\linewidth]{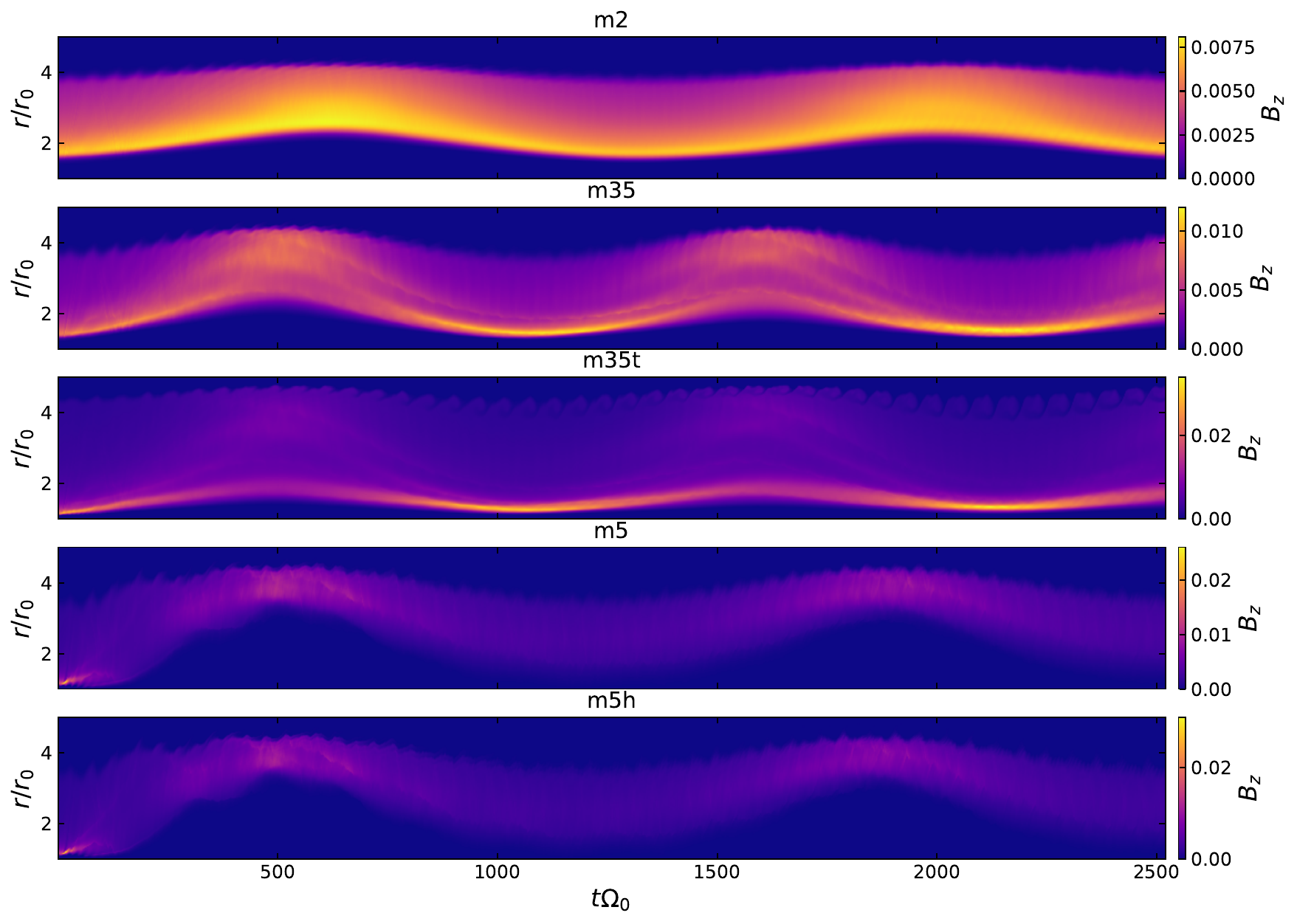}
\caption{Same as Fig. \ref{fig:spcvr}, but for vertical magnetic field in the MHD simulations. }
\label{fig:spcbz}
\end{figure*}

For a given eccentricity profile and resolution, our hydrodynamic and MHD simulations show remarkably similar evolution. Table \ref{tab:sims} compares the precession frequencies we observe in simulations against the eigenvalues computed in generating our initial conditions. We measure precession rates in the simulations by fitting lines to binned arguments of pericentre $\tilde{\omega}$ that have been averaged over the interior of the disc. The measured precession frequencies agree reasonably well with the predicted eigenvalues, \textit{except} in simulations that we identify as under-resolved (namely m35l, h5, m5, m5h). For the simulations, we also estimate eccentricity decay rates (listed as imaginary parts of the frequencies) by fitting slopes to the natural logarithm of the AMD as a function of time, and assuming that the AMD decays as $e^2.$

The spacetime diagrams in Fig. \ref{fig:spcvr} show radial profiles of radial velocity as a function of time for most of the hydrodynamic and MHD simulations listed in Table \ref{tab:sims}. Sliced at a fixed $\phi=0,$ these spacetime diagrams illustrate the coherent precession of untwisted eccentric distortions for maximum eccentricities of $0.2$ and $0.35.$ The modes initialised with $\max[e]=0.5$ involve very strong eccentricity gradients (and hence density variations) near the inner boundary, coming closer to the limiting eccentric mode shape for our radial extent $r_1/r_0=5.$ Their interaction with the inner boundary leads to shock formation that is visible in the bottom three panels of Fig. \ref{fig:spcvr}. Although the eccentric distortions in these simulations continue to precess, they clearly take on a different character from the initial conditions. \cite{Papaloizou05b} observed such shocks in simulations initialised with linear eccentric modes prescribed a finite amplitude, and \cite{Barker16} excluded them by considering only smaller values of $\max[e].$ 

For a given value of the maximum eccentricity in the simulation domain, these spacetime diagrams show very little difference between the hydrodynamic and MHD simulations; the vertical magnetic field simply causes slightly more rapid precession. The spacetime diagrams in Fig. \ref{fig:spcbz} illustrate the corresponding evolution of the vertical field with time. The simulations m35 and m35t have similar values of $\max[e],$ but two different widths of ``envelope'' (see Equation \ref{eq:env}) for the vertical magnetic field ($w_{\rm transition}=0.1$ and $0.05,$ respectively). The different distributions of vertical magnetic flux do little to alter the characteristics or behaviour of the eccentric mode. 

\begin{figure}
\includegraphics[width=\linewidth]{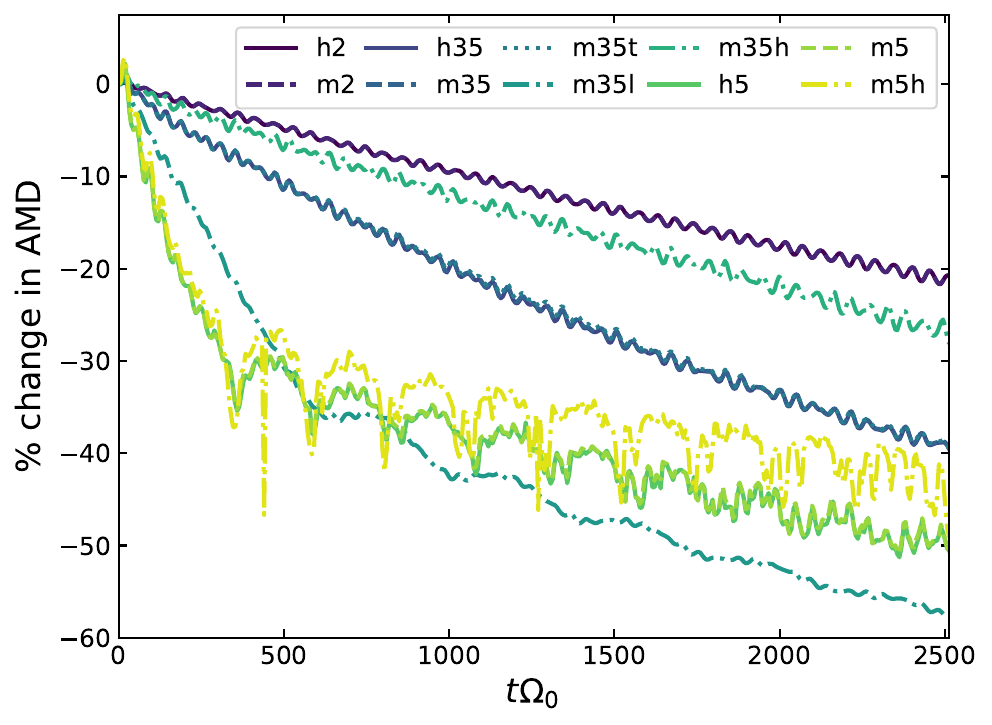}
\caption{Time-evolution of angular momentum deficit (Equation \ref{eq:AMD}) for all of the simulations listed in Table \ref{tab:sims}. The AMD of the magnetised and unmagnetised discs are nearly indistinguishable (e.g. see the curves for h2 and m2; h35 and m35; and h5 and m5).}
\label{fig:amd}
\end{figure}

\begin{figure}
\includegraphics[width=\linewidth]{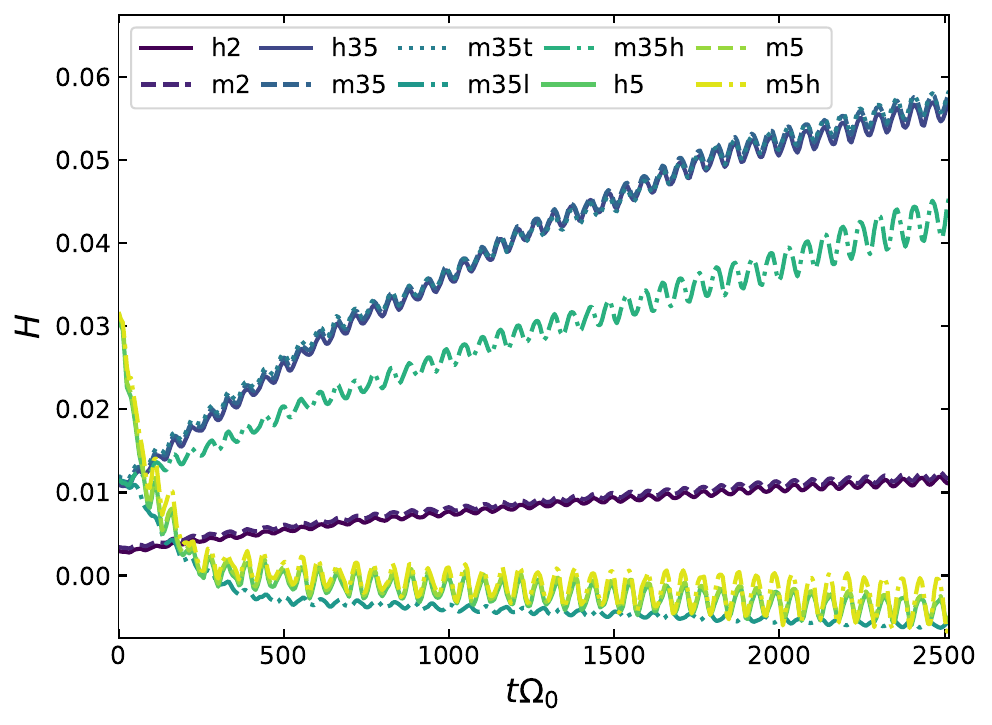}
\caption{Time-evolution of the conserved Hamiltonian (Equation \ref{eq:H}) for all of the simulations listed in Table \ref{tab:sims}. As with Figure \ref{fig:amd} the evolution of the total-Hamiltonian for the magnetised and unmagnetised disc are nearly indistinguishable.}
\label{fig:H}
\end{figure}

\begin{figure}
\includegraphics[width=\linewidth]{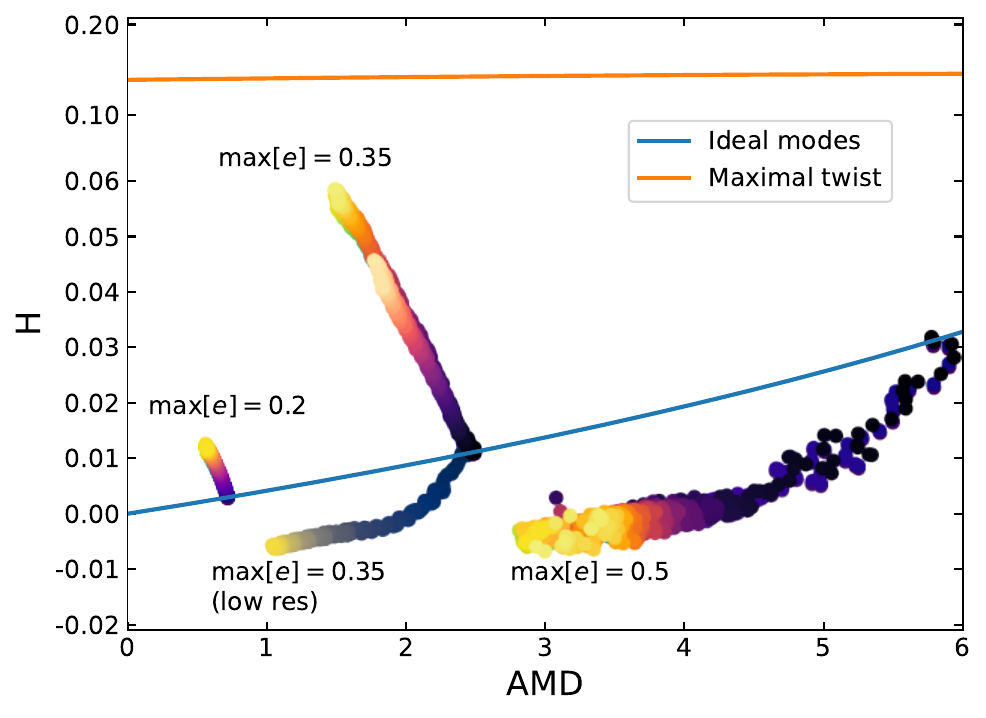}
\caption{Plot showing simulations' time evolution in $\mathcal{H}$-AMD phase space. All ten simulations from Figs. \ref{fig:amd} and \ref{fig:H} are represented by points with colours that run from dark to light as time progresses. The blue and orange curves respectively illustrate the relationship between $\mathcal{H}$ and AMD for ideal untwisted eccentric modes, and eccentric distortions with twists set everywhere equal to the values required for orbital intersection. For better visibility, the y-scale transitions from log to linear at $\mathcal{H}=0.06.$}
\label{fig:HvAMD}
\end{figure}

Fig. \ref{fig:amd} provides a quantitative measure of eccentricity decay, plotting the per cent change in integrated AMD versus time. For lower eccentricities ($\max[e]=0.2,0.35$) we attribute eccentricity decay both to numerical diffusion, and to weak damping by the initial growth of the Papaloizou-Pringle instability \citep[see ][]{Barker16}. The level of decay over the simulation runtimes (of $400T_0,$ where $T_0=2\pi/\Omega_0$ and $\Omega_0=\sqrt{GM/a_\text{in}^3}$ are the orbital period and angular velocity at the inner boundary) is consistent with the hydrodynamic results reported by \cite{Barker16}. 

The shock formation in the simulations with larger $\max[e]=0.5$ leads to much stronger eccentricity damping initially (until $t\Omega_0\simeq400$), and shallower, ``bursty'' decay at later times. We attribute this stochastic evolution to periodic interaction between the strongly modified distortion and the inner boundary. Fig. \ref{fig:amd} quantitatively demonstrates the similarity of the hydrodynamic and MHD results for a given eccentricity profile and resolution, regardless of vertical flux distribution (compare m35 and m35t). 

The curves in Fig. \ref{fig:H} show the evolution of the total Hamiltonian (Equation \ref{eq:H}), which is clearly not conserved in our simulations. Adiabatic damping, i.e. damping which is slow relative to the precession timescale, should lead to a slow evolution of the eccentricity along the family of ideal eccentric modes, towards modes of lower amplitude. As in the unmagnetised case \citep{Ogilvie19}, equation \ref{variational form} implies that an infinitesimal change in the total Hamiltonian is related to an infinitesimal change in the AMD by $d \mathcal{H} = - \omega d \mathcal{C}$. The modes in our simulations have retrograde precession $(\omega<0)$ meaning a decreases in AMD should lead to a decreasing $\mathcal{H}$, when damping is slow enough. Thus slowly damped modes should follow the blue curve in Fig. \ref{fig:HvAMD} which shows the $\mathcal{H}$-AMD phase space. However, Fig. \ref{fig:H}-\ref{fig:HvAMD} demonstrates secular growth for the simulations with $\max[e]=0.2,0.35$ (except for the low-resolution simulation m35l, which shows similar decay to the simulations with $\max[e]=0.5$). 

One potential explanation for this growth is that non-adiabatic damping in our simulations shifts the initialised eccentric profiles away from the family of ideal eccentric modes that minimise $\mathcal{H}$ for a given AMD. In particular when damping is strong enough the eccentric disc will develop a twist as a result of the disc transporting AMD to compensate for spatial variations of the damping rate \citep{Ferreira09}. Fig. \ref{fig:HvAMD} shows that the resolved simulations evolve from the untwisted "modal" $\mathcal{H}$-AMD relation (blue curve) to the "maximally twisted" $\mathcal{H}$-AMD relation (Orange Curve). The latter is obtained by taking a given eccentric mode and twisting it until it reaches an orbital intersection everywhere. This is consistent with the disc gradually twisting, over the course of the simulation, causing a growth in the total Hamiltonian. This is supported by the simulations orbital elements, computed using \ref{eq:e from sim}, and from looking at the residual in the radial velocity when the radial velocity of the untwisted eccentricity profile is subtracted; both of which show the disc becoming increasingly twisted with time. This twisting of the disc occurs over 1000s of orbits and is thus much milder than that seen for non-modal initial profiles \citep[e.g. the const-$e$ profiles studied by][]{Chan22} which become highly twisted over 10s of orbital periods. 

\begin{figure}
\includegraphics[width=\linewidth]{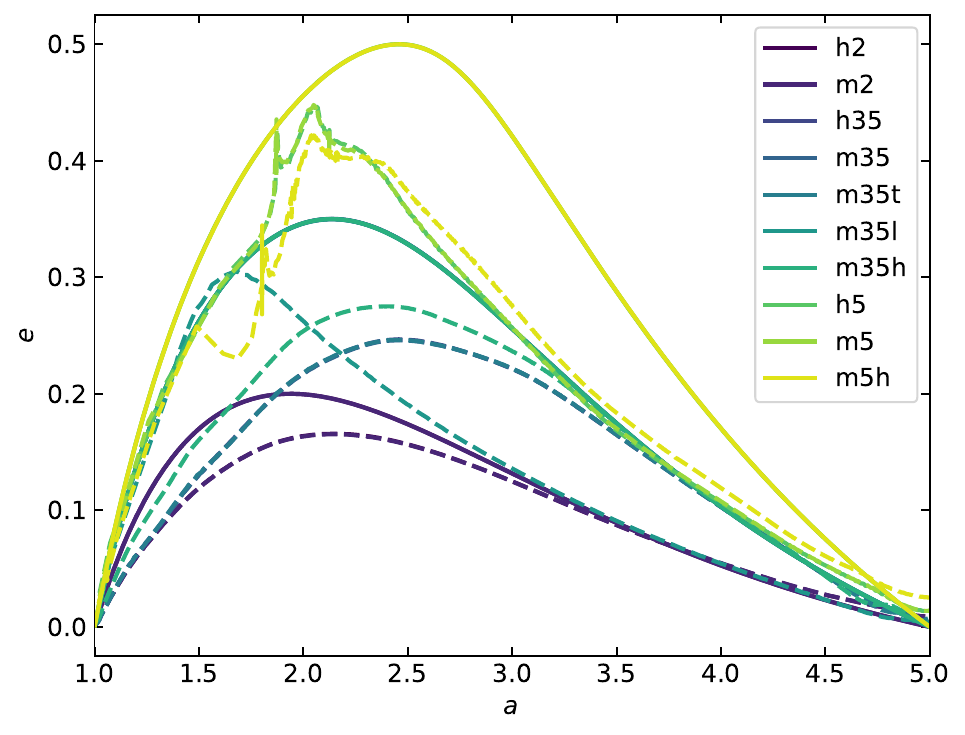}
\caption{Binned eccentricity profiles at the beginning (solid) and end (dashed) of our simulations. The modes with $\max[e]=0.2,0.35$ decay to qualitatively similar eccentricity profiles (except in the low-resolution run m35l). On the other hand, shocks in the simulations initialised with eccentric distortions with $\max[e]=0.5$ lead to strongly modified eccentricity profiles.}
\label{fig:eprof}
\end{figure}

\begin{figure*}
\includegraphics[width=\textwidth]{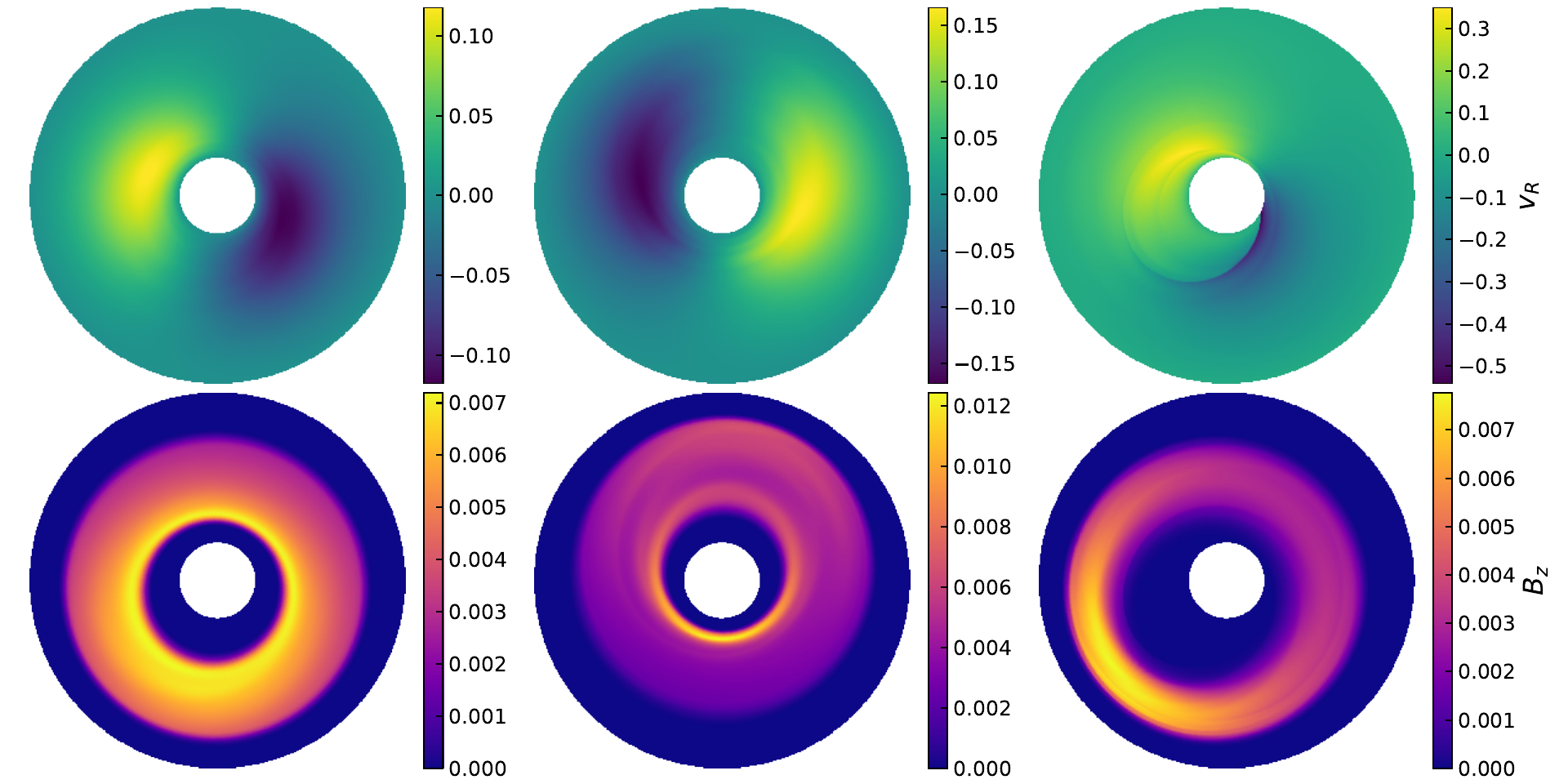}
\caption{Snapshots showing end-state radial velocities (top) and vertical magnetic fields (bottom) in the simulations m2, m35, and m5. Steep density gradients in m5 lead to the development of shocks that are likely unresolved.}\label{fig:snap}
\end{figure*}

Fig. \ref{fig:eprof} plots profiles of binned eccentricity $\tilde{e}(a)$ at the beginning (solid lines) and end (dashed lines) of our simulations. For $\max[e]=0.2$ and $0.35,$ the plot shows the decay of eccentric modes that retain roughly the same profile in eccentricity, except in m35l (which has half the radial and azimuthal resolution). Although the profiles for the simulations with $\max[e]=0.5$ deviate qualitatively, h5, m5 and m5h remain strongly distorted and relatively untwisted by the end of the simulations. The panels in Fig. \ref{fig:snap} show snapshots of radial velocity (top) and vertical magnetic field (bottom) at the end of the simulations m2 (left), m35 (middle), and m5 (right).

The differences with increasing resolution illustrated by Figs. \ref{fig:amd}-\ref{fig:eprof} (compare m35,m35l,m35h, and m5,m5h) indicate that care should be taken in resolving disk distortions with strong eccentricity gradients. We do not claim to have completely resolved the eccentric modes' precession in any of our simulations; m35l, m35, and m35h demonstrate a clear reduction in AMD decay with increasing resolution. However, this decay is slow compared with the dynamical timescales of interest for magnetorotational and parametric instabilities. Further, m35 and m35h exhibit qualitatively similar if not quantitatively identical evolution.

\section{Discussion} \label{discussion}

At the magnetic field strength relevant to the MRI the magnetic field has negligible influence on the eccentric modes, which are almost indistinguishable from their unmagnetised counterparts. In 2D (i.e. specifically suppressing the MRI and parametric instability) the evolution of the magnetised and unmagnetised eccentric modes in RAMSES are qualitatively the same. There are some differences seen between the magnetised and unmagnetised simulations with $\max[e]=0.5$, however these simulations are not adequately resolved. 

Although the magnetic field has little effect on the eccentric mode, the presence of an eccentric mode can have a strong influence on the magnetic field: with lateral compression by the orbital motion, the presence of eccentricity gradients can enhance the magnetic field strength in regions of the disc. Similarly the magnetic field strength is reduced in regions of the disc where the orbital velocity diverges. Despite the strong magnetic field enhancements, the magnetic field configurations we setup are stable in our 2D simulations and their slow evolution is consistent with that expected due to the evolution of the eccentricity profile. In the simulated modes the enhancement of the magnetic field is primarily a result of the imposition of circular rigid wall boundaries. However, this effect is potentially very important in short wavelength/tightly-wound eccentric discs such as those expected in the inner regions of black hole discs as simulated by \citet{Dewberry20}.

One issue that we have encountered is the difficulty of both resolving and converging the eccentric modes in numerical simulations. This is important for studies of the eccentric MRI, as having a high enough resolution to resolve the MRI (e.g. as measured by MRI quality factors) may not be sufficient to ensure the simulation is well resolved. One also needs adequate horizontal resolution to resolve the the eccentric mode as well. This is particularly important if one is interested in analysing the effects of the MRI on the eccentric disc as the strong damping of the eccentricity by the grid may overwhelm the effects of the MHD turbulence. Given the relatively strong damping seen in our 2D simulations assessing the influence of the MRI on the eccentric disc will prove challenging unless MRI is very efficient at damping (or in principle exciting) eccentricity.

In this paper we have limited our focus to the 2.5D cylindrical disc setup. This setup has a number of advantages numerically (easier to implement the vertical boundary and to achieve adequate vertical resolution), however it does not give a good approximation to a physical 3D disc. As in hydrodynamic eccentric discs the variation of vertical gravity and pressure around an eccentric orbit leads to a dynamically varying scale height around an orbit. This causes prograde precession of the eccentric disc \citep{Ogilvie01,Ogilvie08,Ogilvie14,Ogilvie19}. More importantly the vertical compression induced by the scale height oscillation can greatly enhance the quasi-toroidal magnetic fields in nonlinearly eccentric discs \citep{Lynch21}. Additionally the periodic solution to the induction equation within the disc needs to match onto the current free external field. This can be constructed in a circular disc via matched asymptotics \citep{Ogilvie97}. However for non-axisymmetric discs the set of external field solutions (which can be described using cylindrical harmonics) are generically incompatible with the field configuration within an eccentric disc (excepting the purely quasi-toroidal case where no magnetic flux leaves the disc). The internal and external fields could be connected by a force free transition layer in the upper disc atmosphere. However such a field configuration is likely unstable even in the absence of the MRI.

The full 3D problem is important, however, and deserves further attention. A simpler initial approach might be to simulate a 3D MHD disc while exciting eccentricity at the outer boundary \citep[Similar to][]{Dewberry20} and observe the magnetic field response.

\section{Conclusion} \label{conclusion}

In this paper we have extended the Hamiltonian eccentric disc theory of \citet{Ogilvie19} to include a magnetic field in an unstratified, cylindrical geometry. We have solved for the uniformly precessing eccentric mode solutions of our model and shown that, for magnetic field strengths relevant to the onset of MRI, the resulting eccentricity profile, and precession rate, is nearly identical to the unmagnetised case. While such eccentric modes are of limited utility in describing realistic 3D eccentric discs due to several important physical effects not being present in the unstratified geometry, they provide a useful setting for the study of the eccentric MRI and magnetised parametric instability to further our understanding of how disc turbulence operates in eccentric discs. More broadly, this will help inform our understanding of how disc turbulence operates in flows that vary on the orbital timescale. To this end we confirm the suitability of our eccentric mode solutions for numerical applications by using them as initial conditions for 2D MHD simulations in RAMSES. In 2D simulations we obtain long lived uniformly precessing eccentric flows that agree closely with the analytical predictions. These flows will provide the background state for 3D, unstratified, simulations studying the stability of these eccentric discs to both the MRI and parametric instability which will be presented in a future publication.

\section*{Acknowledgements}
 The authors would like to thank Guillaume Laibe and Enrico Ragusa for many helpful comments on the draft of this manuscript and the anonymous reviewer for comments and suggestions, which improved the clarity of the paper.
 
 E. Lynch would like to thank the European Research Council (ERC). This research was supported by the ERC through the CoG project PODCAST No 864965. This project has received funding from the European Union’s Horizon 2020 research and innovation program under the Marie Skłodowska-Curie grant agreement No 823823. 

 J. Dewberry gratefully acknowledges support from the Natural
Sciences and Engineering Research Council of Canada (NSERC), [funding
reference \#CITA 490888-16].

\section*{Data availability}

The data underlying this article will be shared on reasonable request to the corresponding author.



\bibliographystyle{mnras}
\bibliography{MHD_eccentric_sims}




\onecolumn

\appendix

\section{Modes with free boundaries} \label{free boundaries}

As discussed in \citet{Ogilvie19} the large eccentricity gradients seen in our simulated modes, which are responsible for the strong variation of the density around the orbit, are primarily a consequence of imposing rigid circular boundaries. More realistic free boundaries, appropriate for an eccentric disc of finite extent, tend to result in smaller eccentricity gradients for a given value of $\max[e]$, reducing the level of magnetic field concentration by the eccentric mode.

Following the same procedure laid out in \citet{Zanazzi20}, one obtains the free boundary conditions by introducing a taper $T(a)$\footnote{Note $T(a)\ne W(a)$ the taper in the magnetic field as the latter is designed to isolate the magnetic field from the boundaries}, which drops to zero on the disc boundary, into the disc mass and internal energy by taking $m_a \rightarrow m_a T(a)$ and $\langle \bar{\varepsilon} \rangle \rightarrow \langle \bar{\varepsilon} \rangle T(a)$. In an isothermal disc the latter implies the sound speed also drops to zero on the boundary with $c_s^2 \propto T(a)$. This is in fact a requirement for the disc to truncate without the forces due to pressure gradients exceeding those due to gravity and dominating the dynamics of the fluid in the outer disc. Taking the lengthscale of the taper to zero we require

\begin{equation}
 \frac{\partial F_{V_t, V_z}}{\partial f} \Biggl|_{a_{\rm min}, a_{\rm max}} = 0 ,
 \label{free bc}
\end{equation}
in order that the precession frequency remain finite.

In the presence of a taper in the magnetic field (e.g. the magnetic fields given by Equations \ref{vz case}-\ref{vt case}) then Equation \ref{free bc} simplifies to

\begin{equation}
 \frac{\partial F^{(\gamma)}}{\partial f} \Biggl|_{a_{\rm min}, a_{\rm max}} = 0 ,
 \label{free bc hydro}
\end{equation}
which is matches the free boundary condition in the unmagnetised disc.

We now consider vertical field setup for the modes computed in Section \ref{nonlinear modes}, but impose Equation \ref{free bc hydro} for the free boundaries. This results in the modes depicted in Figure \ref{free bc} (left). Such mode have non-zero eccentricity on both boundaries and are thus a challenge to simulate numerically. They do, however, posses the shallower eccentricity gradients and monotonically decreasing profiles characteristic of the fundamental mode in more realistic setups. As shown by Figure \ref{free bc} (right) the variation of the plasma-$\beta$ around the orbit is much milder than that seen in the rigid wall case and suggests the fundamental mode in eccentric disc does not strongly enhance the magnetic field over the circular value. As discussed in Section \ref{nonlinear modes}, this is not the case for higher order modes which can support larger eccentricity gradients, and thus magnetic field enhancements, independently of which boundary condition is adopted.

\begin{figure}
\begin{subfigure}{0.5\textwidth}
\includegraphics[width=\linewidth]{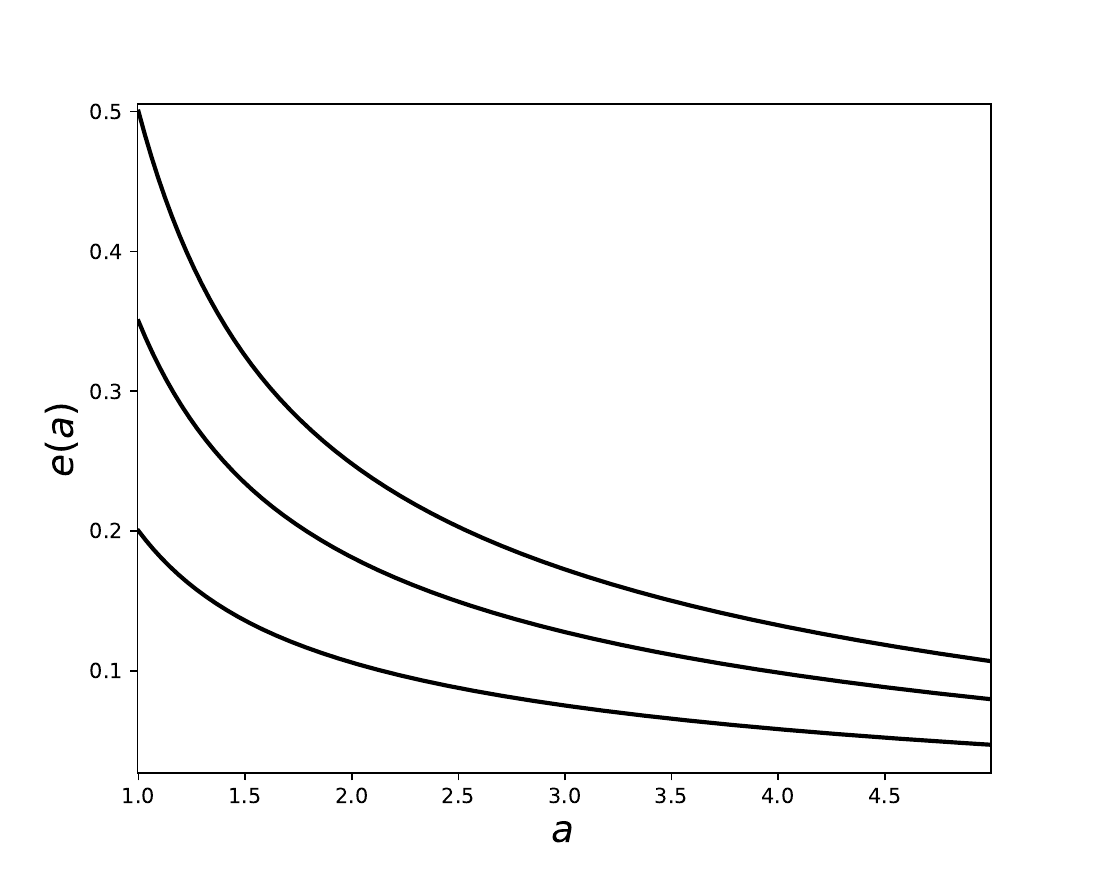}
\label{free bc}
\end{subfigure}
\begin{subfigure}{0.5\textwidth}
\includegraphics[width=\linewidth]{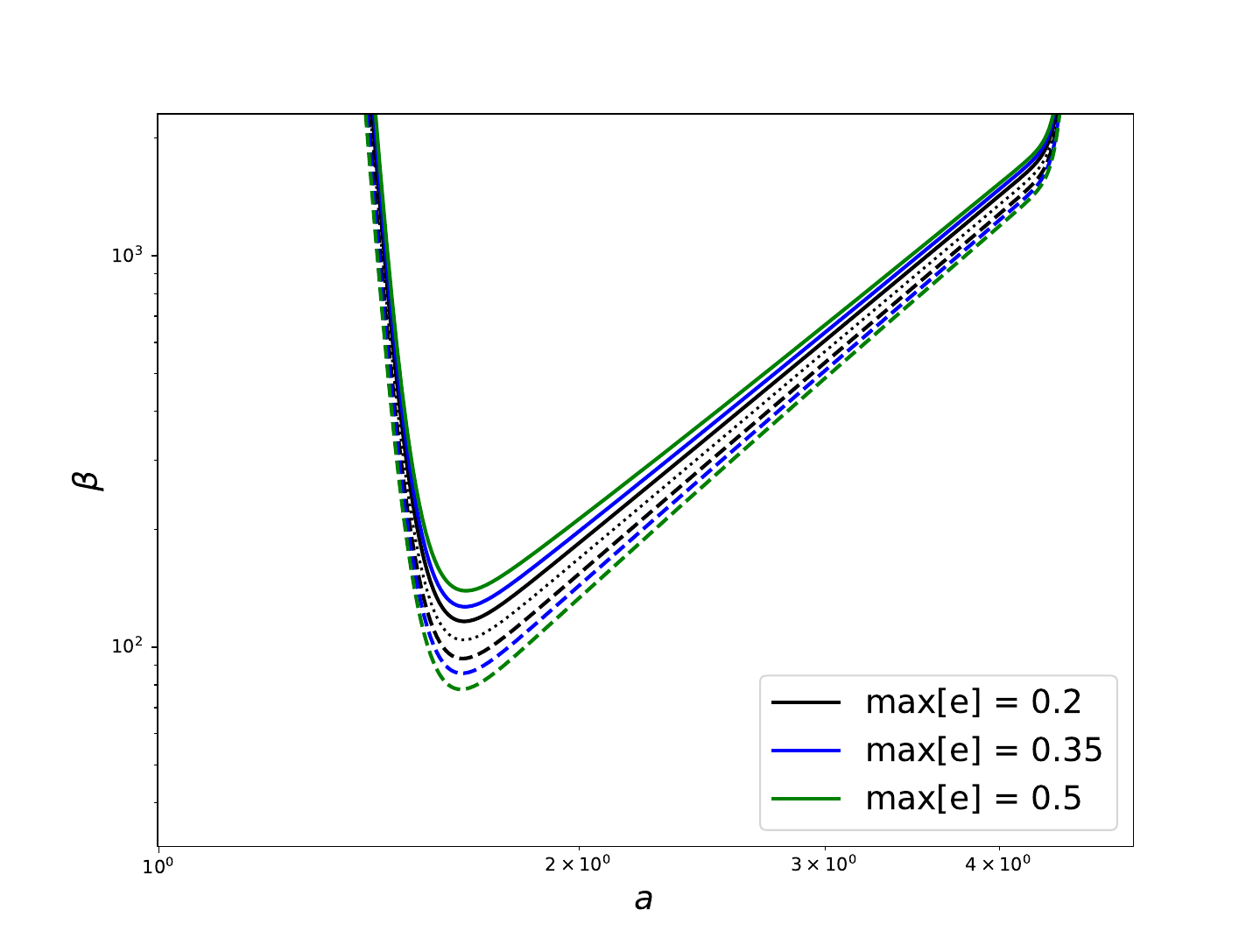}
\label{plasma beta free}
\end{subfigure}
\caption{(Left) Example eccentric modes with free boundaries. (Right) Same as Figure \ref{plasma beta bz} but for the modes with free boundaries. The equivalent modes with quasi-toroidal fields look broadly similar. There is much less enhancement of the magnetic field as modes with realistic boundaries are less steep than those with rigid boundaries.}
\end{figure}

\section{Input variables for numerical simulations}

\subsection{Input Variables for the RAMSES simulations}

In the $a,\phi$ grid the input variables required for simulations, for a given eccentricity profile, $e(a)$, are

\begin{align}\label{eq:rho_from_e}
 \rho &= \rho^{\circ} \left( \frac{1 - e (e + a e_a)}{\sqrt{1 - e^2}} - \frac{a e_a}{\sqrt{1 - e^2}} \frac{\cos \phi + e}{1 + e \cos \phi} \right)^{-1}\\
 u_R &= \frac{a n}{\sqrt{1 - e^2}} e \sin \phi \\
 u_{\phi} &= \frac{a n}{\sqrt{1 - e^2}} (1 + e \cos \phi)\\\label{eq:Bz_from_e}
 B_z &= \frac{n l_{\rm MRI}}{2 \pi \sqrt{16/15}} W (a) \left( \frac{1 - e (e + a e_a)}{\sqrt{1 - e^2}} - \frac{a e_a}{\sqrt{1 - e^2}} \frac{\cos \phi + e}{1 + e \cos \phi} \right)^{-1}
\end{align}
These are then interpolated onto the cylindrical grid by performing a 1D interpolation from $a$ to $R$ for each azimuthal slice in the ($a$, $\phi$) grid and copying the resulting 2D fields in the vertical direction (if present).
 

\subsection{Vector Potential} \label{appendix vector pontential}

For numerical implementations it is often useful to specify the magnetic field using a vector potential in order to ensure that the magnetic field is divergence free. For example transforming from the orbital coordinate system to cylindrical polars can induce a non-zero velocity divergence if interpolation is done on the B-field. Instead interpolating the A-field ensures the resulting magnetic field, in cylindrical polars, obeys the solenoidal condition.

The vector potential obeys

\begin{equation}
\dot{A}_i - \epsilon_{i j k} u^{j} \epsilon^{k a b} \partial_a A_{b} = \partial_{i} f
 \label{A field evo}
\end{equation}
where an overdot indicates a partial derivative with respect to time, $\epsilon_{i j k}$ indicates the permutation symbol and $f$ is the electrostatic potential. We are interested in magnetic fields which are steady on the orbital timescale, for such magnetic fields Equation \ref{A field evo} simplifies to

\begin{equation}
2 u^{j} \partial_{[j} A_{i]} = \partial_{i} f , 
 \label{A field evo steady}
\end{equation}
where the square brackets around the indices denote that the indices are being anti-symmetrised, i.e. $T_{[i j]} = \frac{1}{2} (T_{i j} - T_{j i})$. As we did when computing the $B$-field, we neglect the contribution from the slow rotation of the field due to disc precession.

In the $(a,M,\tilde{z})$ orbital coordinate system, with stretched vertical coordinate $\tilde{z}=z/H$, the fluid velocity simplifies to $u^{i} = n \hat{e}^{i}_{M}$, where $n$ is the mean motion. The steady $B$-field in a eccentric disc, in the $(a,M,\tilde{z})$ orbital coordinate system is

\begin{equation}
 B^{i} = B_{t 0} (a , \tilde{z}) j^{-1} h^{-1} \hat{e}_{\tilde{M}}^{i} + B_{z 0} (a) j^{-1} H^{-1} \hat{e}_{\tilde{z}}^{i} .
\end{equation}
Using the relationship between the $A$ and $B$ fields we find that the $A$ field must obey the following expressions in order to yield the steady magnetic field solution in an eccentric disc,

\begin{align}
 0 &= \partial_{[M} A_{\tilde{z} ]} , \label{M z component} \\
 \frac{B_{t 0} (a , \tilde{z})}{ j h } &= - \frac{2}{J H} \partial_{[a} A_{\tilde{z}]}  , \\
 \frac{B_{z 0} (a)}{j H} &= \frac{2}{J H} \partial_{[a} A_{M]} .
\end{align}
Substituting the expression for the velocity and Equation \ref{M z component} into Equation \ref{A field evo steady} we find that the electrostatic potential is a function of semimajor axis only, $f = f(a)$. This means the $M$ and $\tilde{z}$ components of Equation \ref{A field evo steady} are satisfied. To obtain agreement with the $B$ field solution we set $A_{a} = 0$ and obtain the following for $A_{M}$ and $A_{\tilde{z}}$,

\begin{align}
 A_{M} &= \int J^{\circ} B_{z 0} (a) \, d a , \\
 A_{\tilde{z}} &= -\int H^{\circ} J^{\circ} B_{t 0} (a , \tilde{z}) \, d a .
\end{align}
In order to satisfy the $a$ component of Equation \ref{A field evo steady} we require the electrostatic potential satisfy

\begin{equation}
 f (a) = - \int n J^{\circ} B_{z 0} \, d a .
\end{equation}
We now transform to the $(a,E,z)$ orbital coordinate system, which is more useful for transforming $\boldsymbol{A}$ into other coordinate systems. To do this we assume that we have a thin disc, allowing us to evaluate the coordinate transform at the midplane similar to \citet{Ogilvie18}. This works equally well for the unstratified discs considered in the rest of this paper. The vector potential in this coordinate system is

\begin{align}
 A_{a} &= - e_a \sin E \int J^{\circ} B_{z 0} (a) \, d a , \\
 A_{E} &= (1 - e \cos E) \int J^{\circ} B_{z 0} (a) \, d a , \\
 A_{z} &= -\frac{1}{H} \int H^{\circ} J^{\circ} B_{t 0} (a , \tilde{z}) \, d a .
\end{align}

The general expression for $\boldsymbol{A}$ in Cartesian and cylindrical coordinates are fairly complicated, but can be obtained from the above expressions as follows

\begin{align}
A_{x} &= -\frac{1}{J} \left( e_a \sin E \frac{\partial y}{\partial E} + (1 - e \cos E) \frac{\partial y}{\partial a}  \right) \int J^{\circ} B_{z 0} (a) \, d a , \\
A_{y} &= \frac{1}{J} \left( e_a \sin E \frac{\partial x}{\partial E} + (1 - e \cos E) \frac{\partial x}{\partial a}  \right) \int J^{\circ} B_{z 0} (a) \, d a , \\
A_{z} &= -\frac{1}{H} \int H^{\circ} J^{\circ} B_{t 0} (a , \tilde{z}) \, d a ,
\end{align}
while the expression in cylindrical polars can be obtained in the usual manor from $A_{R} = \cos \phi A_x + \sin \phi A_y$ and $A_{\phi} = -r \sin \phi A_x + r \cos \phi A_y$. To evaluate these we require the following expressions for the partial derivatives of the Cartesian coordinates,

\begin{align}
\frac{\partial x}{\partial a} &= (\cos E - e) \cos \varpi - \frac{1 - e (e + a e_a)}{\sqrt{1 - e^2}} \sin E \sin \varpi - a e_a \cos \varpi - a (\cos E - e) \varpi_a \sin \varpi - a \varpi_a \sqrt{1 - e^2} \sin E \cos \varpi , \\
\frac{\partial x}{\partial E} &= -a \sin E \cos \varpi - a \sqrt{1 - e^2} \cos E \sin \varpi , \\
\frac{\partial y}{\partial a} &= \frac{1 - e (e + a e_a)}{\sqrt{1 - e^2}} \sin E \cos \varpi + (\cos E - e) \sin \varpi - a e_a \sin \varpi - a \varpi_a \sqrt{1 - e^2} \sin E \sin \varpi + a \varpi_a (\cos E - e) \cos \varpi , \\
\frac{\partial y}{\partial E} &= a \sqrt{1 - e^2} \cos E \cos \varpi - a \sin E \sin \varpi .
\end{align}
For an untwisted disc where the pericentre direction is aligned with the $x$-axis these expressions simplify significantly and we arrive at the following expressions for $\boldsymbol{A}$ in Cartesian coordinates

\begin{align}
A_{x} &= -\frac{\sin E}{J} \left( a e_a  \sqrt{1 - e^2} \cos E + \frac{1 - e (e + a e_a)}{\sqrt{1 - e^2}}  (1 - e \cos E) \right) \int J^{\circ} B_{z 0} (a) \, d a , \\
A_{y} &= \frac{1}{J} \left( -a e_a \sin^2 E +  (\cos E - e - a e_a)  (1 - e \cos E) \right) \int J^{\circ} B_{z 0} (a) \, d a , \\
A_{z} &= -\frac{1}{H} \int H^{\circ} J^{\circ} B_{t 0} (a , \tilde{z}) \, d a ,
\end{align}
while in cylindrical polars $\boldsymbol{A}$ is given by

\begin{align}
A_{R} &= \frac{\sin E}{J} \left( -\cos \phi a e_a  \sqrt{1 - e^2} \cos E  -\cos \phi \frac{a e_a}{\sqrt{1 - e^2}}  (1 - e \cos E) - \sin \phi a e_a \sin E - a e_a \sqrt{1 - e^2}  \right) \int J^{\circ} B_{z 0} (a) \, d a , \\
\begin{split}
A_{\phi} &= \frac{r}{J} \left( \sin E \sin \phi a e_a  \sqrt{1 - e^2} \cos E + \sin \phi \sin E \frac{1 - e (e + a e_a)}{\sqrt{1 - e^2}}  (1 - e \cos E) -a e_a \cos \phi \sin^2 E +  \cos \phi (\cos E - e - a e_a)  (1 - e \cos E)  \right) \\
&\times \int J^{\circ} B_{z 0} (a) \, d a , 
\end{split} \\
A_{z} &= -\frac{1}{H} \int H^{\circ} J^{\circ} B_{t 0} (a , \tilde{z}) \, d a .
\end{align}


\bsp	
\label{lastpage}
\end{document}